\documentclass[aps,prb,reprint,twocolumn, longbibliography]{revtex4-2}
\pdfoutput=1

\usepackage{graphicx}
\usepackage[space]{grffile}
\usepackage{amsmath,amsfonts,amssymb,mathtools}
\usepackage{color}
\usepackage{subfigure}
\usepackage{dsfont}
\usepackage{hyperref}
\usepackage{booktabs}
\usepackage{url}
\usepackage{leftidx}
\usepackage{textcomp}
\usepackage{gensymb}
\usepackage{threeparttable}
\usepackage{rotating}
\usepackage{esvect}
\usepackage{booktabs}
\usepackage{lmodern}
\usepackage[noend]{algpseudocode}
\usepackage[dvipsnames]{xcolor}
\usepackage{adjustbox}
\usepackage{multirow}
\usepackage{svg}
\usepackage[normalem]{ulem}

\graphicspath{{../Inkscape/},{../Mathematica/},{./Figures/},{./images/}}

\begin{document}

\title{Unconventional superconductivity protected from disorder on the kagome lattice}

\author{Sofie Castro Holb{\ae}k}
\affiliation{Niels Bohr Institute, University of Copenhagen, DK-2200 Copenhagen, Denmark}

\author{Morten H. Christensen}
\affiliation{Niels Bohr Institute, University of Copenhagen, DK-2200 Copenhagen, Denmark}

\author{Andreas Kreisel}
\affiliation{Niels Bohr Institute, University of Copenhagen, DK-2200 Copenhagen, Denmark}

\author{Brian M. Andersen}
\affiliation{Niels Bohr Institute, University of Copenhagen, DK-2200 Copenhagen, Denmark}

\date{\today}

\begin{abstract}
Motivated by the recent discovery of superconductivity in the kagome $A$V$_3$Sb$_5$ ($A$: K, Rb, Cs) metals, we perform a theoretical study of the symmetry-allowed superconducting orders on the two-dimensional kagome lattice with focus on their response to disorder. We uncover a qualitative difference between the robustness of intraband spin-singlet (even-parity) and spin-triplet (odd-parity) unconventional superconductivity to atomic-scale nonmagnetic disorder. Due to the particular sublattice character of the electronic states on the kagome lattice, disorder in spin-singlet superconducting phases is only weakly pair-breaking despite the fact that the gap structure features sign changes. By contrast, spin-triplet condensates remain fragile to disorder on the kagome lattice. We demonstrate these effects in terms of the absence of impurity bound states and an associated weak disorder-induced $T_c$-suppression for spin-singlet order. We also discuss the consequences for quasi-particle interference and their inherent tendency for momentum-space anisotropy due to sublattice effects on the kagome lattice. For unconventional kagome superconductors, our results imply that any allowed spin-singlet order, including for example $d+id$-wave superconductivity, exhibits a disorder-response qualitatively similar to standard conventional $s$-wave superconductors.

\end{abstract}

\maketitle

\section{Introduction}

The irreducible representations of a given point group dictate the allowed homogeneous superconducting order parameters of materials~\cite{SigristUeda91}. However, additional complexity of e.g. sublattice degrees of freedom, multiple active orbitals at the Fermi level, associated Hund's interactions, or strong spin-orbit coupling adds significant richness to the problem~\cite{Scalapino2012,Chubukov2012Pairing,Steglich2016Foundations,Leereview,HIRSCHFELD2016197,VafekHund,Kreisel_review,KreiselOS,Fischerreview,Romer2019}. Whatever superconducting gap structure eventually gets singled out in a given material depends on the particular pairing mechanism operative in that system. Determining experimentally which order parameter is present in specific materials can be a tremendous challenge due to the low temperature and often minute energy scales of the superconducting problem. In this respect, disorder can play an important role because it can act as a phase-sensitive probe. Disorder effects can be studied in terms of e.g. atomically-resolved impurity bound states detectable by local probes as well as the overall disorder-averaged superconducting response as seen by e.g. thermodynamic probes or transport measurements. If nonmagnetic disorder is able to generate in-gap bound states or if it severely affects superconductivity, it is typically a strong indicator of an unconventional superconducting condensate~\cite{Balatsky_review}.   

\begin{figure}
\centering
\includegraphics[width=0.9\linewidth]{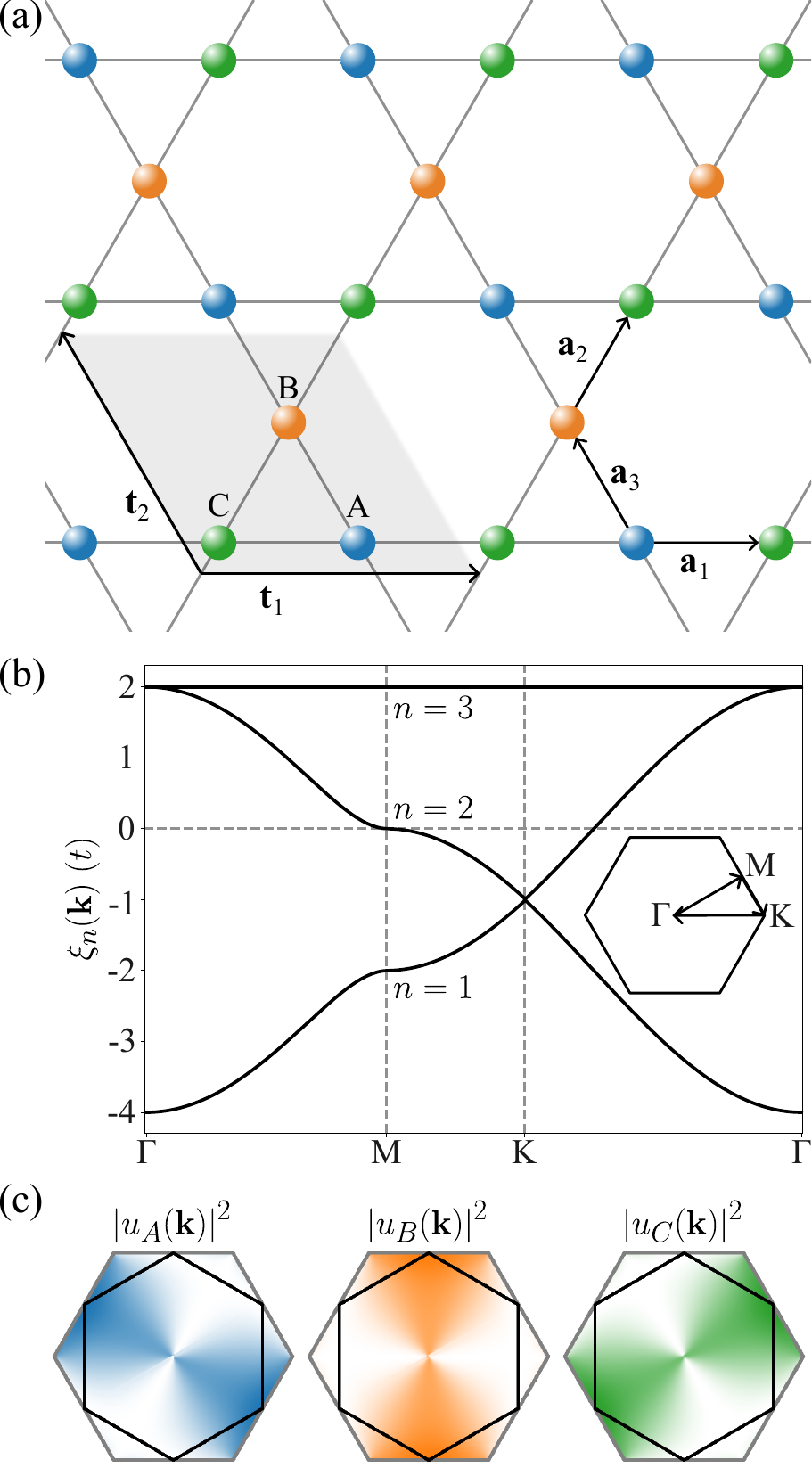}
\caption{\label{fig:kagome_lattice_basics}(a) Illustration of the kagome lattice with lattice vectors $\mathbf{t}_1$ and $\mathbf{t}_2$, and sublattice vectors $\mathbf{a}_1$, $\mathbf{a}_2$ and $\mathbf{a}_3$. The grey area denotes the unit cell with the three different sublattice sites denoted A (blue), B (orange) and C (green). (b) Tight-binding energy bands plotted along the high-symmetry path shown in the inset. Here, $n$ is a band label, introduced for later convenience. (c) Distribution of the sublattice weights across the Brillouin zone for the second band ($n=2$). The Fermi surface for $\mu = 0$, corresponding to the upper van Hove singularity in (b), is shown in black.}
\end{figure}

The discovery of superconductivity in vanadium-based kagome metals $A$V$_3$Sb$_5$ ($A$: K, Rb, Cs) has reinvigorated the discussion of conventional versus unconventional pairing in novel quantum materials~\cite{Ortiz2020CsV3Sb5}. The kagome lattice is particularly intriguing since its basic electronic structure features both flat bands, van Hove singularities, and Dirac points, as seen in Fig.~\ref{fig:kagome_lattice_basics}. In addition, the Fermi surface distribution of sublattice weights of the eigenstates of the kagome tight-binding bands, also illustrated in Fig.~\ref{fig:kagome_lattice_basics}, can play important roles for determining e.g. the leading instabilities arising from electronic interactions~\cite{KieselEA12,Kiesel2013Unconventional}. This may indeed be of relevance in the $A$V$_3$Sb$_5$ compounds where superconductivity appears in proximity to a charge-density wave (CDW) phase, which has been studied intensely both theoretically~\cite{ParkEA21, LinEA22,Denner2021,Tazai2022mechanism, Christensen2021, Ferrari2022,Christensen2022} and experimentally~\cite{Ortiz2019New, Kenney2021Absence, Jiang2021Unconventional, Chen2021Roton, Zhao2021Cascade}. For the superconducting phase, some theoretical studies have explored Cooper pairing arising from purely electronic fluctuations~\cite{Yu2012,KieselEA12,Wang2013,ParkEA21,Wu2021Nature,Tazai2022mechanism,Wen2022superconducting,RomerEA22,Wu2021Nature,He2022Strong-coupling,LinEA22,Bai2022effective}. Other works have pointed to the important role of phonons for the generation of superconductivity, either on their own or in conjunction with electronic correlations~\cite{Wu2EA22,TanEA21,Zhang2021firstprinciples,Zhong2022Testing,Wang_phonon_2023,Ritz23}. 

Experimentally, the $A$V$_3$Sb$_5$ kagome metals enter their superconducting phase at $T_c\sim 1-3$K~\cite{Ortiz2020CsV3Sb5,OrtizEA21,YinEA21}. The critical temperature $T_c$, however, may be significantly enhanced by uniaxial strain or hydrostatic pressure, e.g., for CsV$_3$Sb$_5$, $T_c\sim 8$K at $\sim 2$~GPa~\cite{Qian2021Revealing,Chen2021Double,Gupta2022Two,Du2021Pressure-induced}. The detailed nature and origin of electronic pairing in $A$V$_3$Sb$_5$ remains controversial at present, with evidence for both standard nodeless non-sign-changing gaps and nodal unconventional superconducting order. For example, several STM measurements have reported 'V'-shaped STM conductance spectra~\cite{Chen2021Roton,Liang2021Three-dimensional,Xu2021Multiband}, and thermal conductivity data has been interpreted in favor of nodal superconductivity~\cite{Zhao2021nodal}. Similarly, muon spin spectroscopy experiments on RbV$_3$Sb$_5$ and KV$_3$Sb$_5$ samples report nodal gaps at ambient pressure~\cite{Guguchia2022Tunable}, and a pressure-tuned transition to nodeless order with additional evidence for spontaneous time-reversal symmetry breaking (TRSB) setting in at $T_c$ for high pressures $\sim 2$~GPa~\cite{Mielke2022Time-reversal,Guguchia2022Tunable}.
On the other hand, a Knight shift suppression and the existence of a Hebel-Slichter coherence peak in the spin-lattice relaxation observed by nuclear magnetic resonance measurements point to $s$-wave spin-singlet superconductivity~\cite{Mu2021S-wave}.
Penetration depth measurements and specific heat data on CsV$_3$Sb$_5$ have also been analysed in terms of an anisotropic non-sign-changing gap with a finite small minimum gap~\cite{OrtizEA21,Duan2021Nodeless,Gupta2022Microscopic,RoppongiEA22,Gupta2022Two}. Recent laser ARPES measurements find isotropic (momentum-independent) spectroscopic gaps~\cite{ZhongEA23}. Nodeless superconductivity also appears consistent with multiband features and a 'U'-shaped conductance behavior near zero bias as seen by some scanning tunneling microscopy (STM) measurements~\cite{Xu2021Multiband}. In addition, the lack of sign-changes in the gap function appears consistent with the absence of in-gap states near nonmagnetic impurities~\cite{Xu2021Multiband} and a weak dependence of $T_c$ on residual resistivity ratios (sample quality)~\cite{Zhang2023}.

A main point of the current work is that one needs to be careful applying standard arguments about disorder and pair-breaking when the impurities are located on sites that are not centers of the point group operations. This is indeed the case for the kagome lattice, where the site-symmetry of the individual lattice sites does not possess the full point group symmetry. In a nutshell, even though a gap function averages to zero over the Fermi surface, $\sum_{\bf k}\Delta({\bf{k}})=0$, atomic-scale disorder necessarily selects a specific sublattice site and thereby only scatters to certain allowed regions of the Fermi surface. For spin-singlet order on the kagome lattice this mechanism of sublattice-selective scattering in fact renders nonmagnetic disorder nearly blind to sign changes in the gap structure. In this sense, the kagome lattice protects spin-singlet superconductivity from nonmagnetic disorder. 

Here, we perform a theoretical study of the basic properties of the superconducting phases on the kagome lattice and their response to pointlike defects, and determine the robustness of different symmetry-allowed superconducting order parameters to disorder. As mentioned above, we demonstrate that the sublattice weight of the band eigenstates play a crucial role for determining the response of superconductivity to disorder. Specifically, while intraband spin-triplet pairing remains fragile to disorder, spin-singlet superconductivity is robust to atomic-scale disorder due to the sublattice dependence of the eigenstates. We stress that this is a generic property of the kagome lattice, and not restricted to a specific electron concentration and associated Fermi surface. More generally, we expect that other lattices with substantial sublattice weight variation present at the Fermi surface will exhibit similar behavior. We analyze the impact of atomic scale disorder in terms of allowed impurity bound states and the resulting local density of states, and discuss implications for quasi-particle interference. Additionally, we discuss the robustness of $T_c$ within Abrikosov-Gor'kov theory~\cite{AG_classic} applied to both spin-triplet and spin-singlet superconducting states, and compare to similar calculations obtained on the square lattice.

The paper is organized as follows: In Sec.~\ref{sec:kagome_lattice} we review the electronic structure of the kagome lattice and in Sec.~\ref{sec:kagome_SC} we construct the symmetry-allowed superconducting orders assuming a single $s$-orbital degree of freedom on each sublattice site. From these candidates we next explore their response and robustness to atomic scale disorder. Section~\ref{sec:disorder_kagome} focuses on the existence of impurity bound states and the response to local probes, whereas the disorder-averaged impact on $T_c$ within an Abrikosov-Gor'kov approach is described in Sec.~\ref{sec:AG}. Finally, Sec.~\ref{sec:discussion} contains a discussion of the relevance of the current findings to the $A$V$_3$Sb$_5$ materials, and our conclusions.

\section{Phenomenology of the kagome lattice}\label{sec:kagome_lattice}

The kagome lattice is a network of corner-sharing triangles with three sites in the unit cell, as shown in Fig.~\ref{fig:kagome_lattice_basics}. The normal state Hamiltonian reads
\begin{equation}
    \mathcal{H}_0 = \sum_{\substack{\mathbf{k}\sigma \\ \alpha\beta}} \left(h_{\alpha\beta}(\mathbf{k}) - \mu\delta_{\alpha\beta} \right)c^{\dagger}_{\mathbf{k}\sigma\alpha}c^{\phantom{\dagger}}_{\mathbf{k}\sigma\beta}\,,
\end{equation}
where $\sigma$ denotes spin and $\alpha,\beta=A, B, C$ are sublattice indices. The kinetic part is given by
\begin{align}
    h(\mathbf{k}) =
    -2t\begin{pmatrix}
        0   &     \cos k_3    &   \cos k_1 \\
        \cos k_3    &   0    &   \cos k_2 \\
        \cos k_1     &    \cos k_2 &    0
    \end{pmatrix}\,,
\end{align}
where we have restricted to nearest-neighbor (NN) hoppings. In what follows we will set the hopping amplitude $t=1$. Here, $k_i = \mathbf{k} \cdot \mathbf{a}_i$ where
\begin{equation}
    \mathbf{a}_1 = \frac{1}{2}\begin{pmatrix}
        1 & 0
    \end{pmatrix}\,, \quad 
    \mathbf{a}_2 = \frac{1}{2}\begin{pmatrix}
        \frac{1}{2} & \frac{\sqrt{3}}{2}
    \end{pmatrix}\,, \quad 
    \mathbf{a}_3 = \frac{1}{2}\begin{pmatrix}
        -\frac{1}{2} & \frac{\sqrt{3}}{2}
    \end{pmatrix}\,,
\end{equation}
are the sublattice vectors, where we have put the distance between neighboring unit cells to one.

The Hamiltonian is diagonalized by a unitary transformation, $u_{n \alpha}^{\ast}(\mathbf{k}) h_{\alpha\beta}(\mathbf{k}) u_{\beta m}(\mathbf{k})=\xi_n(\mathbf{k}) \delta_{nm}$ yielding the band energies $\xi_n(\mathbf{k})$ and the eigenstates $u_{\alpha n}(\mathbf{k})$ of band $n$.
The resulting band structure is shown in Fig.~\ref{fig:kagome_lattice_basics} and features a Dirac point at the K point and two van Hove singularities at the M point. In addition, for the NN tight-binding model there is a flat band which acquires a weak dispersion upon including further neighbor hoppings. Furthermore, the kagome lattice is endowed with a property which has become known as sublattice interference~\cite{KieselEA12} for which specific hopping trajectories interfere destructively resulting in electronic wavefunctions that localize on specific sites inside the unit cell. Specifically, electronic states at the upper van Hove singularity, at $\mu=0$, are localized on only one of the three sites in the unit cell, as shown in Fig.~\ref{fig:kagome_lattice_basics}. In contrast, states at the lower van Hove singularity, at $\mu=-2$, localize on two of the three sublattice sites.

\section{Superconducting pairing states}\label{sec:kagome_SC}

Superconductivity is included at the mean-field level through the Bogoliubov-de Gennes (BdG) framework. In Nambu basis, the Hamiltonian reads 
\begin{equation}
    \mathcal{H}_{\rm BdG} = \sum_{\mathbf{k}} \Psi^{\dagger}_{\mathbf{k}}
    \widehat{H}_{\rm BdG}(\mathbf{k})
    \Psi_{\mathbf{k}}\,,
\end{equation}
where 
\begin{equation}
    \widehat{H}_{\rm BdG}(\mathbf{k}) = \begin{pmatrix}
        h(\mathbf{k}) - \mu\mathds{1} & -\Delta(\mathbf{k})\\
        - \Delta(\mathbf{k})^{\dagger} & -h(-\mathbf{k})^T + \mu\mathds{1}
    \end{pmatrix}\,,\label{eq:BdG_ham}
\end{equation}
and $\Psi_{\mathbf{k}}^{\dagger} = \begin{pmatrix} c^{\dagger}_{\mathbf{k} \uparrow} & c^{\phantom{\dagger}}_{-\mathbf{k} \downarrow} \end{pmatrix}$ with $c^{\dagger}_{\mathbf{k}\sigma}=\begin{pmatrix}c^{\dagger}_{\mathbf{k}\sigma A}&c^{\dagger}_{\mathbf{k}\sigma B}&c^{\dagger}_{\mathbf{k}\sigma C}\end{pmatrix}$ and the hat denotes $6 \times 6$ matrices in Nambu and sublattice space. In this section we discuss which form the $3 \times 3$ matrices $\Delta(\mathbf{k})$ can take on the kagome lattice. Readers interested mainly in the effects of disorder can skip directly to Sec.~\ref{sec:disorder_general}. In what follows, we will not be concerned with the microscopic origin of superconductivity and instead focus on which states are allowed to exist by symmetry. A momentum independent microscopic pairing interaction induces only on-site terms while any momentum dependence of the pairing interaction generically yields both on-site terms and further neighbor couplings.  As shown in Fig.~\ref{fig:kagome_lattice_basics}, the two-dimensional (2D) kagome lattice has three distinct sites in the unit cell and the point group is $D_6$. Barring any accidental near-degeneracies, the superconducting order parameter will therefore transform as an irreducible representation (irrep) of $D_6$.

\begin{table}[!tb]
\resizebox{\columnwidth}{!}{
\begin{tabular}{ccccc}
\toprule
\multirow{2}{*}{\textbf{Irrep}} & \multicolumn{2}{c}{\textbf{Real space}} & \multicolumn{2}{c}{\textbf{Momentum space}}\\
& \textbf{Singlet} & \textbf{Triplet} & \textbf{Singlet} & \textbf{Triplet}\\
\hline
On-site & & & & \\
$A_1$ &  \adjustbox{raise=-0.5\height}{\resizebox{!}{1.5cm}{\includegraphics{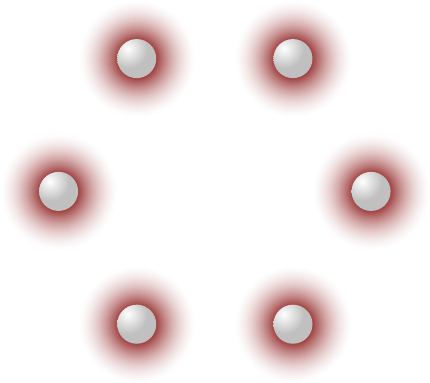}}} & &  \adjustbox{raise=-0.5\height}{\resizebox{!}{1.5cm}{\includegraphics{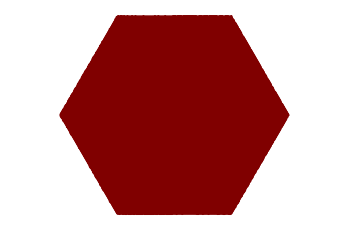}}} &\\
$E_2^{(1)}$ & \adjustbox{raise=-0.5\height}{\resizebox{!}{1.5cm}{\includegraphics{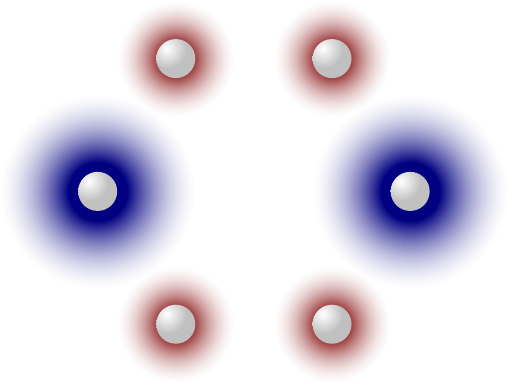}}} &  & \adjustbox{raise=-0.5\height}{\resizebox{!}{1.5cm}{\includegraphics{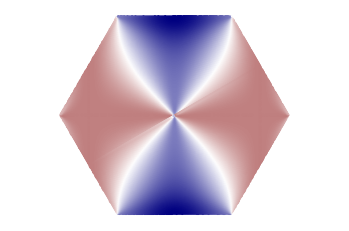}}} &\\
$E_2^{(2)}$ &  \adjustbox{raise=-0.5\height}{\resizebox{!}{1.5cm}{\includegraphics{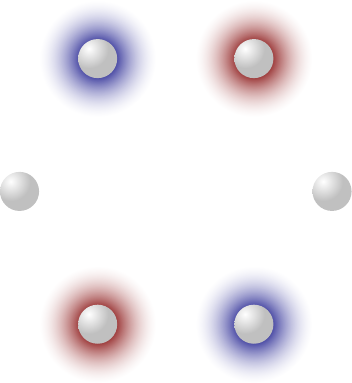}}} & & \adjustbox{raise=-0.5\height}{\resizebox{!}{1.5cm}{\includegraphics{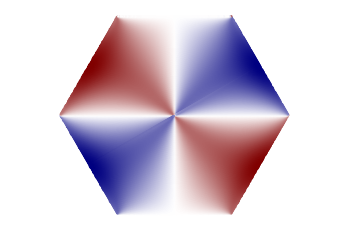}}} &\\
\hline
\!Nearest-neighbor\!\!\!\!\!\!\!\!\!\!\!\!& & & & \\
$A_1$ & \adjustbox{raise=-0.5\height}{\resizebox{!}{1.5cm}{\includegraphics{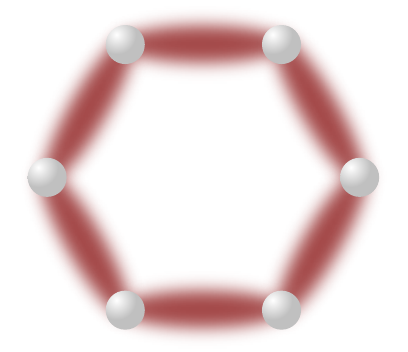}}} & & \adjustbox{raise=-0.5\height}{\resizebox{!}{1.5cm}{\includegraphics{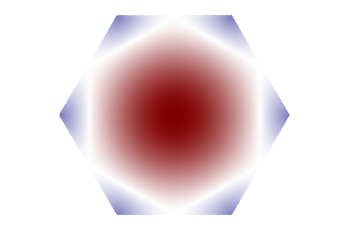}}} &\\
$A_2$ & & \adjustbox{raise=-0.5\height}{\resizebox{!}{1.5cm}{\includegraphics{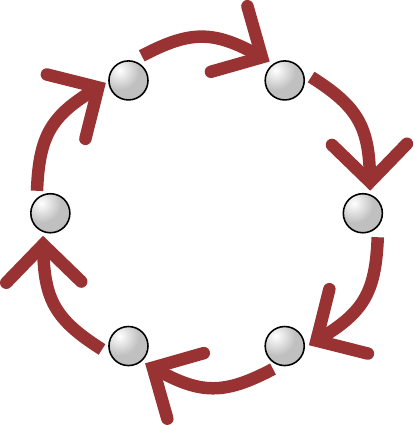}}} & & Interband\\
$B_1$ & & \adjustbox{raise=-0.5\height}{\resizebox{!}{1.5cm}{\includegraphics{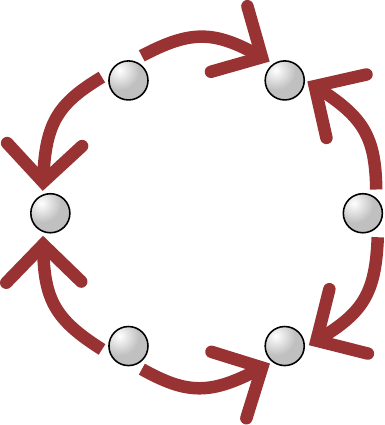}}} & & \adjustbox{raise=-0.5\height}{\resizebox{!}{1.5cm}{\includegraphics{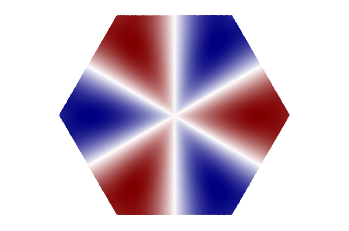}}}\\
$B_2$ & \adjustbox{raise=-0.5\height}{\resizebox{!}{1.5cm}{\includegraphics{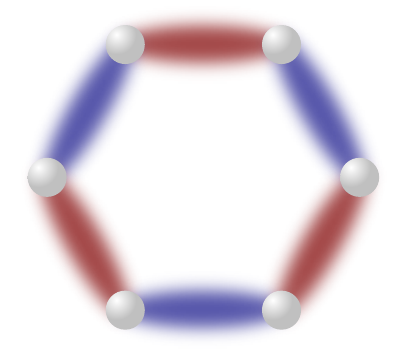}}} & & Interband &\\
$E_1^{(1)}$ & \adjustbox{raise=-0.5\height}{\resizebox{!}{1.5cm}{\includegraphics{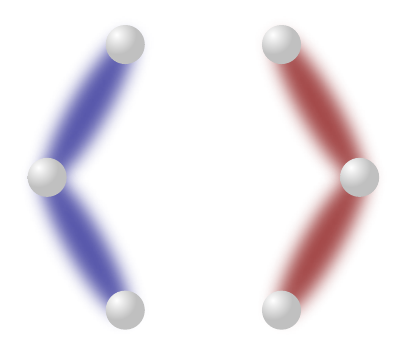}}} & \adjustbox{raise=-0.5\height}{\resizebox{!}{1.75cm}{\includegraphics{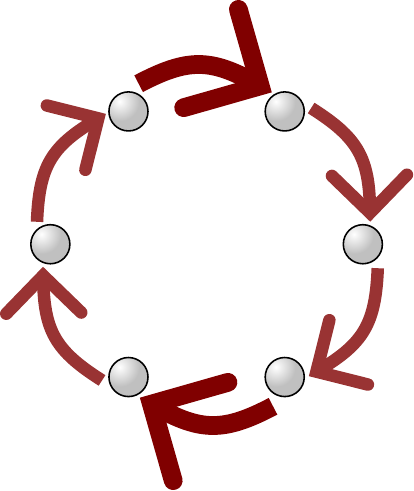}}} & Interband & \adjustbox{raise=-0.5\height}{\resizebox{!}{1.5cm}{\includegraphics{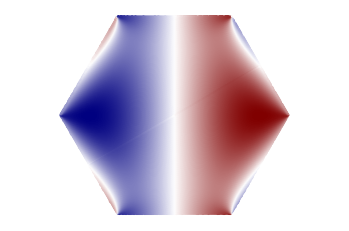}}}\\
$E_1^{(2)}$ & \adjustbox{raise=-0.5\height}{\resizebox{!}{1.5cm}{\includegraphics{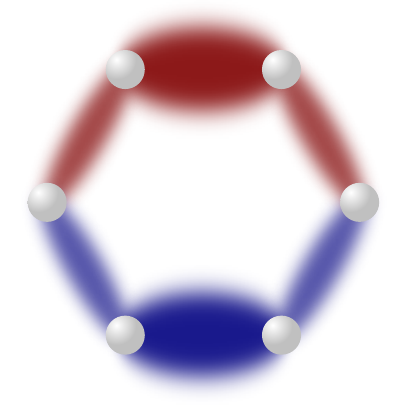}}} & \adjustbox{raise=-0.5\height}{\resizebox{!}{1.25cm}{\includegraphics{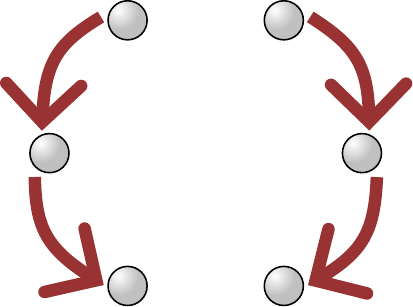}}} & Interband & \adjustbox{raise=-0.5\height}{\resizebox{!}{1.5cm}{\includegraphics{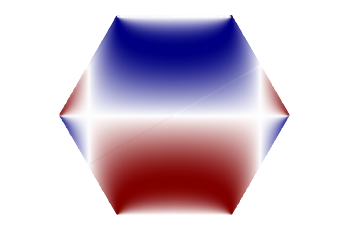}}}\\
$E_2^{(1)}$ & \adjustbox{raise=-0.5\height}{\resizebox{!}{1.5cm}{\includegraphics{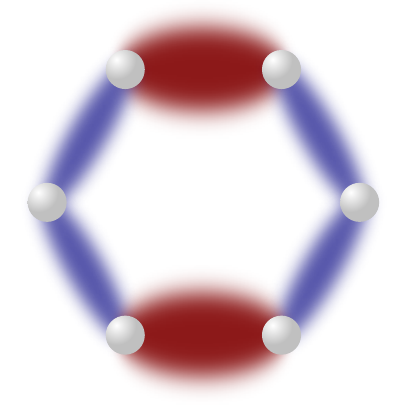}}} & \adjustbox{raise=-0.5\height}{\resizebox{!}{1.25cm}{\includegraphics{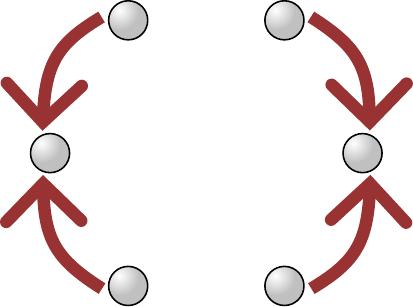}}} & \adjustbox{raise=-0.5\height}{\resizebox{!}{1.5cm}{\includegraphics{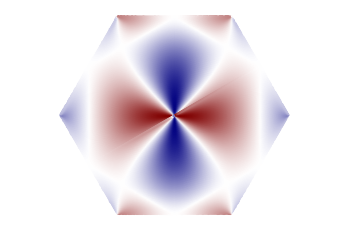}}} & Interband\\
$E_2^{(2)}$ & \adjustbox{raise=-0.5\height}{\resizebox{!}{1.5cm}{\includegraphics{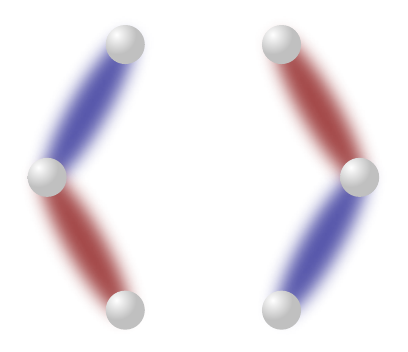}}} & \adjustbox{raise=-0.5\height}{\resizebox{!}{1.75cm}{\includegraphics{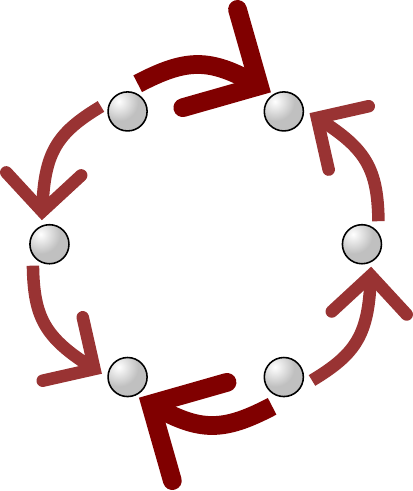}}} & \adjustbox{raise=-0.5\height}{\resizebox{!}{1.5cm}{\includegraphics{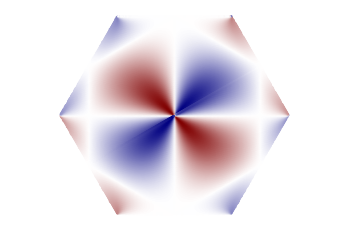}}} & Interband\\
\bottomrule
\end{tabular}}
\caption{\label{tab:SC_OP}Summary of the different pairing states on the kagome lattice. The first column gives the irrep, while the second and third columns provide schematic illustrations of the form of the pairing state in real space. The fourth and fifth columns display the projection of the form factors on the $n=2$ band, as obtained from Eq.~\eqref{eq:gap_transformation}. The irreps on the upper three rows originate from on-site terms while the rest are from nearest-neighbor terms. The color and its intensity denotes the relative sign and strength of the pairing condensate in both the real and momentum space cases. In the triplet case, the sign is denoted by the direction of the arrows, to highlight their antisymmetry under exchange of the real-space indices.}
\end{table}

There are four one-dimensional (1D) irreps, $A_1$, $A_2$, $B_1$, and $B_2$, and two 2D ones, $E_1$ and $E_2$. In what follows, we will decompose the possible pairing states in terms of these irreps. For simplicity, we will confine our attention to pairing states for which the two electrons of the Cooper pair are either on the same site or on neighboring sites. For on-site (OS) Cooper pairs, the pairing state can be decomposed as
\begin{equation}
    f^S_{\rm OS} = A_1 \oplus E_2\,.\label{eq:f_os_singlet}
\end{equation}
The real- and momentum-space structures of these states are shown in Table~\ref{tab:SC_OP}. Accordingly, the two matrices corresponding to the two on-site components of $E_2$ are
\begin{align}
    f^S_{{\rm OS},E_2^{(1)}} &= \frac{1}{\sqrt{6}}\begin{pmatrix}
        +1 & 0 & 0 \\
        0 & -2 & 0 \\
        0 & 0 & +1
    \end{pmatrix}\,, \\
    f^S_{{\rm OS},E_2^{(2)}} &= \frac{1}{\sqrt{2}}\begin{pmatrix}
        +1 & 0 & 0 \\
        0 & 0 & 0 \\
        0 & 0 & -1
    \end{pmatrix}\,.
\end{align}
Since there is only a single orbital per site in our model, on-site states are restricted to being of spin-singlet character. This is no longer the case for pairing states involving NN, and we find both singlet and triplet candidates:
\begin{align}
    f^S_{\rm NN} &= A_1 \oplus B_2 \oplus E_1 \oplus E_2, \label{eq:f_nn_singlet} \\
    f^T_{\rm NN} &= A_2 \oplus B_1 \oplus E_1 \oplus E_2\,, \label{eq:f_nn_triplet}
\end{align}
which are illustrated in Table~\ref{tab:SC_OP}. For example, the two components of the singlet NN $E_2$ form factor written in momentum space are 
\begin{align}
    f^S_{{\rm NN},E_2^{(1)}} &= \frac{1}{\sqrt{12}}\begin{pmatrix}
        0 & - \cos k_3 & 2\cos k_1 \\
        - \cos k_3 & 0 & - \cos k_2 \\
        2 \cos k_1 & -\cos k_2 & 0
    \end{pmatrix}\,, \\
    f^S_{{\rm NN},E_2^{(2)}} &= \frac{1}{2}\begin{pmatrix}
        0 & \cos k_3 & 0 \\
        \cos k_3 & 0 & - \cos k_2 \\
        0 & - \cos k_2 & 0
    \end{pmatrix}\,.
\end{align}
We illustrate the triplet order parameters using arrows to highlight the fact that these, by their nature, are antisymmetric under exchange of the real-space indices. We obtain the triplet $E_1$ components
\begin{align}
    f^T_{{\rm NN},E_1^{(1)}} &= \frac{i}{\sqrt{12}}\begin{pmatrix}
        0 & \sin k_3 & -2 \sin k_1 \\
        \sin k_3 & 0 & -\sin k_2 \\
        -2\sin k_1 & -\sin k_2 & 0
    \end{pmatrix}\,, \\
    f^T_{{\rm NN},E_1^{(2)}} &= \frac{i}{2}\begin{pmatrix}
        0 & \sin k_3 & 0 \\
        \sin k_3 & 0 & \sin k_2 \\
        0 & \sin k_2 & 0
    \end{pmatrix}\,.
\end{align}
The form factors for all the cases shown in Table~\ref{tab:SC_OP} are provided in Appendix~\ref{app:PairingMatrices}. We note that the appearance of the $B_2$ and $E_1$ singlet states and the $A_2$ and $E_2$ triplet states is facilitated only by the internal sublattice structure of the kagome lattice. Consequently, these lead to \emph{inter}-band pairing in momentum space~\cite{Christos2023Nodal}. For the filling factors we consider, none of these lead to a gap opening at the Fermi level, and the electronic structure of the Bogoliubov quasiparticles in this energy range is unchanged from the normal state. Consequently, we do not discuss these states further in what follows. The $B_2$ triplet and $A_2$ singlet gap structures familiar from previous studies of superconductivity on the kagome lattice~\cite{Wu2021Nature,RomerEA22} are obtained at second- and third-nearest neighbors, respectively.

The gap structure is generally a sum of multiple terms. For a generic irrep, $\Gamma$, the singlet or triplet order parameter is
\begin{equation}
    \Delta_{\Gamma} = \Delta_0 f_{{\rm OS},\Gamma} + \Delta_1 f_{{\rm NN},\Gamma} + \cdots \,,\label{eq:gap_exp}
\end{equation}
where $\Delta_i$ denote constants, possibly zero, which depend on the specific irrep and pairing interaction considered and $\cdots$ denotes terms beyond NN which we do not consider in this work. Here, we choose $\Delta_0 = 0.2$ and $\Delta_1 = \Delta_0 /2$ for the $A_1$ and $E_2$ cases and, in the remaining cases, $\Delta_0=0$, $\Delta_1=0.2$. These choices imply that the superconducting coherence peaks are at comparable energies for the different cases. For the 2D irreps, $E_1$ and $E_2$, the system will choose a linear combination of the two components, $a E_{1,2}^{(1)} + b E_{1,2}^{(2)}$, where $a$ and $b$ are complex numbers. Below, we will focus on the cases where $a=1$, $b=i$, which breaks time-reversal symmetry, and $a=1$, $b=0$ (or $a=0$, $b=1$) which breaks lattice rotational symmetry. $\Delta_{\Gamma}$ of Eq.~\eqref{eq:gap_exp} corresponds to the $3\times 3$ matrices appearing in the BdG Hamiltonian, Eq.~\eqref{eq:BdG_ham}. These can be transformed to band space through a unitary transformation
\begin{equation}
    \Delta_{nm}(\mathbf{k}) = u_{n\alpha}^{\ast}(\mathbf{k}) \Delta_{\alpha\beta}(\mathbf{k}) u_{\beta m}^{\ast}(-\mathbf{k})\,, \label{eq:gap_transformation}
\end{equation}
where $u_{n\alpha}(\mathbf{k})$ are the eigenstates of band $n$. We note that $\Delta_{nm}(\mathbf{k})$ is not necessarily a diagonal matrix; the off-diagonal components correspond to interband pairing. The results for the middle band ($n=m=2$) are shown in the fourth and fifth column of Table~\ref{tab:SC_OP}. We note that the projection onto the $n=m=1$ band leads to similar results, as expected from the symmetry of the lattice. The `Interband' cases in Table~\ref{tab:SC_OP} denote the ones where the order parameter consists of electrons from two different bands, i.e. where the off-diagonal elements of $\Delta_{nm}(\mathbf{k})$ are finite while the diagonal components vanish. These states do not generally lead to a gap opening at the Fermi level as mentioned above.

\begin{figure}[tb]
\centering
\includegraphics[width=\linewidth]{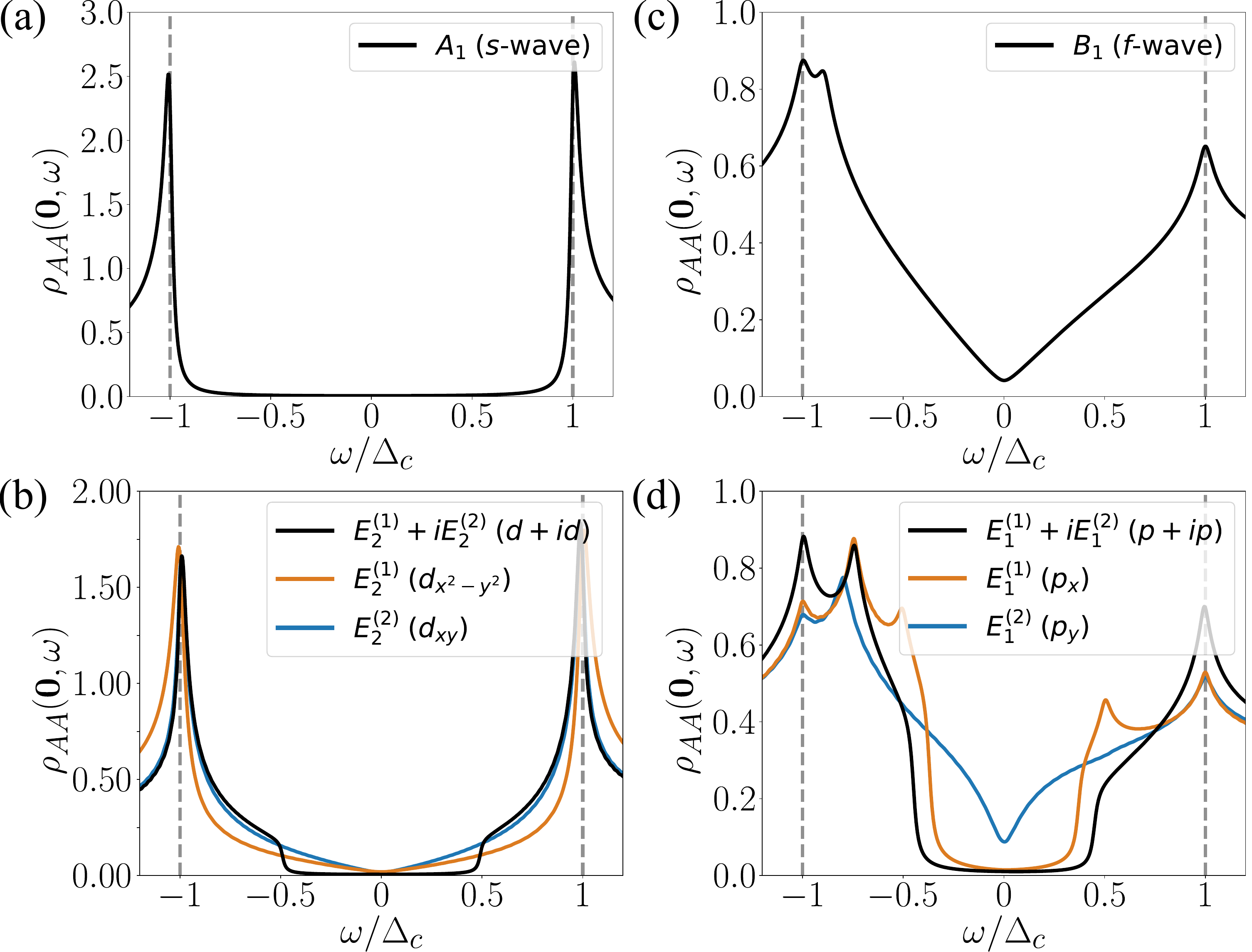}
\caption{\label{fig:clean_SC_DOS}DOS near the Fermi energy for systems with either singlet (left column) or triplet (right column) superconducting order. The singlet cases are plotted for $\mu=0$ while the triplet cases are plotted for $\mu=0.08$. 
(a) Singlet $A_1$, resulting in the usual fully gapped $s$-wave spectrum. (b) Singlet $E_2$ which is 2D and corresponds to $d$-wave at lowest order in momentum. In black we illustrate the $d + id$ TRSB combination. This combination results in a fully gapped spectrum. In orange and blue we show the results for the two individual components, $d_{x^2-y^2}$ (orange) and $d_{xy}$ (blue). Both exhibit nodes on the Fermi surface and lead to V-shaped gaps. They also break rotational symmetry and the LDOS differs between the different sublattice sites. Here we show the case corresponding to the $A$ sublattice. (c) Triplet $B_1$, which is commonly denoted $f$-wave. This has nodes on the Fermi surface (see Table~\ref{tab:SC_OP}) resulting in a V-shaped gap. (d) Triplet $E_1$ which is also 2D and corresponds to $p$-wave at lowest order in momentum. We show the TRSB $p+ip$ variant in black, while the $p_x$ and $p_y$ components are shown in orange and blue, respectively. Both $p_x$ and $p_y$ break lattice rotational symmetry. Here, we stress that we only plot the DOS on the $A$ sublattice which is why the $p_x$ component has a U-shaped gap; the $A$ sublattice has vanishing spectral weight in the regions of the BZ where $p_x$ has nodes, compare Fig.~\ref{fig:kagome_lattice_basics}(c) and Table~\ref{tab:SC_OP}.}
\end{figure}

In Fig.~\ref{fig:clean_SC_DOS} we show the density of states (DOS) for different choices of the superconducting order parameter corresponding to the singlets $A_1$ ($s$-wave) and $E_2$ ($d+id$) and the triplets $E_1$ ($p+ip$) and $B_1$ ($f$). The $s$-wave, $d+id$ and $p+ip$ cases all exhibit a full gap whereas the $f$-wave state is nodal. We denote the location of the superconducting coherence peak by $\Delta_c$. For $d+id$ and $p+ip$ pairing, the full gap is facilitated by the breaking of time-reversal symmetry, which allows the system to avoid possible gap nodes on the Fermi surface. For the $d+id$ and the $p+ip$ cases we also include the DOS for the individual components. In this case, rotational symmetry is broken by the superconducting order, and the DOS on the $A$, $B$, and $C$ sublattices can be different. The cases shown in Fig.~\ref{fig:clean_SC_DOS} correspond to the local density of states (LDOS) at an $A$ sublattice site.

\section{Effects of disorder on kagome superconductivity}
\label{sec:disorder_general}
In this section we turn to a discussion of the effects of disorder on the symmetry-distinct pairing states introduced above. The discussion is divided into two parts. First, we address the response of the system to a single impurity and the possibility of localized in-gap bound states and second, we discuss the disorder-averaged suppression of $T_c$ versus the impurity scattering rate.  

\subsection{Impurity bound states and sublattice interference} \label{sec:disorder_kagome}

In this section, we discuss the effects of a single impurity on superconductivity on the kagome lattice. We focus on isolated lattice site-centered defects, and refer to the discussion section for a more general discussion of the effects of other kinds of disorder. In the following, the existence or absence of bound states and the resulting LDOS is studied within the $T$-matrix framework. In this methodology, bound state solutions are analyzed in terms of the existence of poles in the $T$-matrix. These poles can occur at real or complex energies $\omega$ inside the superconducting gap, referring to bound or resonant impurity states, respectively. Generally, sign-changing gap structures support in-gap bound or resonant states from nonmagnetic disorder as opposed to standard $s$-wave non-sign-changing superconductivity which forbids such solutions~\cite{ANDERSON195926,Balatsky_review}

In order to investigate the response of the pairing candidates to single impurities on the kagome lattice, we introduce either a nonmagnetic impurity or a (classical) magnetic impurity. We arbitrarily locate the impurity at a specific atomic site of the lattice. For concreteness, in the following we choose sublattice site $A$ located at $\mathbf{r}=\mathbf{0}$ to host the impurity. The impurity Hamiltonian matrices in Nambu formalism are then given by
\begin{align}
\label{imppotentialp}
    \widehat{H}_{\text{imp}} = V\tau^z \otimes \begin{pmatrix}
        1&0&0\\
        0&0&0\\
        0&0&0
    \end{pmatrix},
\end{align}
for the potential scatterer, and
\begin{align}
\label{imppotentialm}
    \widehat{H}_{\text{imp}} = S_z \tau^0 \otimes \begin{pmatrix}
        1&0&0\\
        0&0&0\\
        0&0&0
    \end{pmatrix},
\end{align}
for the case of a magnetic impurity. Here, $\tau^z$ and $\tau^0$ denote Pauli matrices in Nambu space. The full Green function described by the Hamiltonian including the impurity potential can be obtained within the $T$-matrix approach:
\begin{align}
\widehat{G}(\mathbf{r},\mathbf{r}',\omega) = \widehat{G}^{(0)}(\mathbf{r}-\mathbf{r}',\omega) + \widehat{G}^{(0)}(\mathbf{r},\omega)\widehat{T}(\omega)\widehat{G}^{(0)}(-\mathbf{r}',\omega),
\end{align}
with the $T$-matrix defined by $\widehat{T}(\omega)\equiv \widehat{D}^{-1}(\omega)\widehat{H}_{\rm imp}$ where
\begin{align}
\widehat{D}(\omega) \equiv \left[ \mathds{1} - \widehat{H}_{\rm imp} \widehat{G}^{(0)}(\mathbf{0},\omega) \right].
\end{align}
Here, the free real-space Green function is calculated from the retarded Green function
\begin{equation}
    \widehat{G}^{(0)}(\mathbf{k},\omega) = \left[ (\omega + i\eta)\mathds{1} - \widehat{H}_{\rm BdG}(\mathbf{k})\right]^{-1}\,, \label{eq:Green_func_def}
\end{equation}
by the Fourier transform $\widehat{G}^{(0)}(\mathbf{r},\omega)= \frac{1}{N} \sum_{\mathbf k} \widehat{G}^{(0)}(\mathbf{k},\omega)e^{i{\mathbf k}\cdot{\mathbf r}}$ where $N$ is number of points in momentum space. The parameter $\eta$ in Eq.~\eqref{eq:Green_func_def} is an infinitesimal positive smearing resulting from the analytic continuation to the real frequency axis. The spin-summed electronic LDOS $\rho_\alpha({\mathbf{r}},\omega)$ at sublattice position $\alpha$ is obtained by
\begin{align}
    \rho_\alpha(\mathbf{r},\omega) = -\frac{1}{\pi}\text{Im}\left[ G_{\alpha\alpha}(\mathbf{r},\mathbf{r},\omega) + G_{\bar{\alpha}\bar{\alpha}}(\mathbf{r},\mathbf{r},-\omega) \right],
\end{align}
where $\bar \alpha$ refers to the same sublattice site as $\alpha$ but in the complementary Nambu block. Throughout, we will use system sizes of at least $N_{k_x} \times N_{k_y} = 1500 \times 1500$ to calculate the LDOS and values of the smearing $\eta \leq 0.0025$. As mentioned above, the existence and location of possible impurity bound states is typically analyzed by studying the poles of the $T$-matrix. In the current situation, exemplified with a single impurity on a sublattice $A$ site, we search for solutions of $\mbox{det} [\widehat{D}(\omega)]=0$, which, for a nonmagnetic scattering potential, yields
\begin{widetext}
\begin{align}
\label{BSconditionp}
V^2\left[G^{(0)}_{AA}(\omega) G^{(0)}_{\bar{A}\bar{A}}(\omega) - G^{(0)}_{A\bar A}(\omega) G^{(0)}_{\bar{A}A}(\omega)\right] +V\left[G^{(0)}_{AA}(\omega)-G^{(0)}_{\bar{A}\bar{A}}(\omega)\right]  -1 =0\,,
\end{align}
and 
\begin{align}
\label{BSconditionm}
S_z^2\left[G^{(0)}_{AA}(\omega) G^{(0)}_{\bar{A}\bar{A}}(\omega) - G^{(0)}_{A\bar A}(\omega) G^{(0)}_{\bar{A}A}(\omega)\right] -S_z\left[G^{(0)}_{AA}(\omega)+G^{(0)}_{\bar{A}\bar{A}}(\omega)\right]  +1 =0 \,,
\end{align}
\end{widetext}
for a magnetic scattering potential. Here, the notation refers to momentum-summed Green functions, i.e. $G^{(0)}_{\alpha\beta}(\omega) \equiv G^{(0)}_{\alpha\beta}(\mathbf{0},\omega) = \frac{1}{N} \sum_{\mathbf{k}} G^{(0)}_{\alpha\beta}({\mathbf{k}},\omega)$. Further investigation of these equations requires determination of the functional form of these local free Green function entries. 

As a short intermezzo, we briefly summarize the physics of impurity bound states in standard one-band superconductors. In that case, Eqs.~(\ref{BSconditionp}) and (\ref{BSconditionm}) are unchanged except for the absence of the sublattice degree of freedom. For a conventional $s$-wave superconductor with an isotropic full gap, Eqs.~(\ref{BSconditionp}) and (\ref{BSconditionm}) allow for bound state solutions only for the case of magnetic impurities as discovered by Yu, Shiba, and Rusinov~\cite{Yu,Shiba,Rusinov}. In contrast, no in-gap solutions exist for nonmagnetic potentials in agreement with Anderson's theorem~\cite{ANDERSON195926}. This can be easily verified by utilizing analytical expressions for the momentum-summed diagonal component
\begin{align}
    G^{(0)}_{11}(\omega)= -\frac{\pi \rho(0) \omega}{\sqrt{|\Delta|^2-\omega^2}}\,,
\end{align}
and the off-diagonal (anomalous) component
\begin{align}
\label{swaveoffdiagonal}
   G^{(0)}_{1\bar{1}}(\omega)= \frac{\pi \rho(0) \Delta}{\sqrt{|\Delta|^2-\omega^2}},
\end{align}
of the Green function at energies inside the gap, $\omega<|\Delta|$. Utilizing Eqs.~(\ref{BSconditionp}) and (\ref{BSconditionm}) for determining the existence of bound states, one sees that the properties of the anomalous Green function are crucial. Typically, for sign-preserving pairing such as standard $s$-wave superconductivity $\sum_{\mathbf{k}} \Delta({\mathbf{k}})\neq 0$, in agreement with Eq.~(\ref{swaveoffdiagonal}), disallowing bound state solutions. On the other hand, for unconventional pairing where $\sum_{\mathbf{k}} \Delta({\mathbf{k}})=0$, the anomalous momentum-summed Nambu components vanish, and bound state solutions may exist for both magnetic and nonmagnetic impurities. In the case of nodal gaps, e.g. $d_{x^2-y^2}$-wave pairing as in the cuprates, these states acquire a lifetime and the energies of the associated resonant or virtual states move off the real axis~\cite{Rosengren,Balatsky_review}. 

\begin{figure}[bt]
\centering
\includegraphics[width=\linewidth]{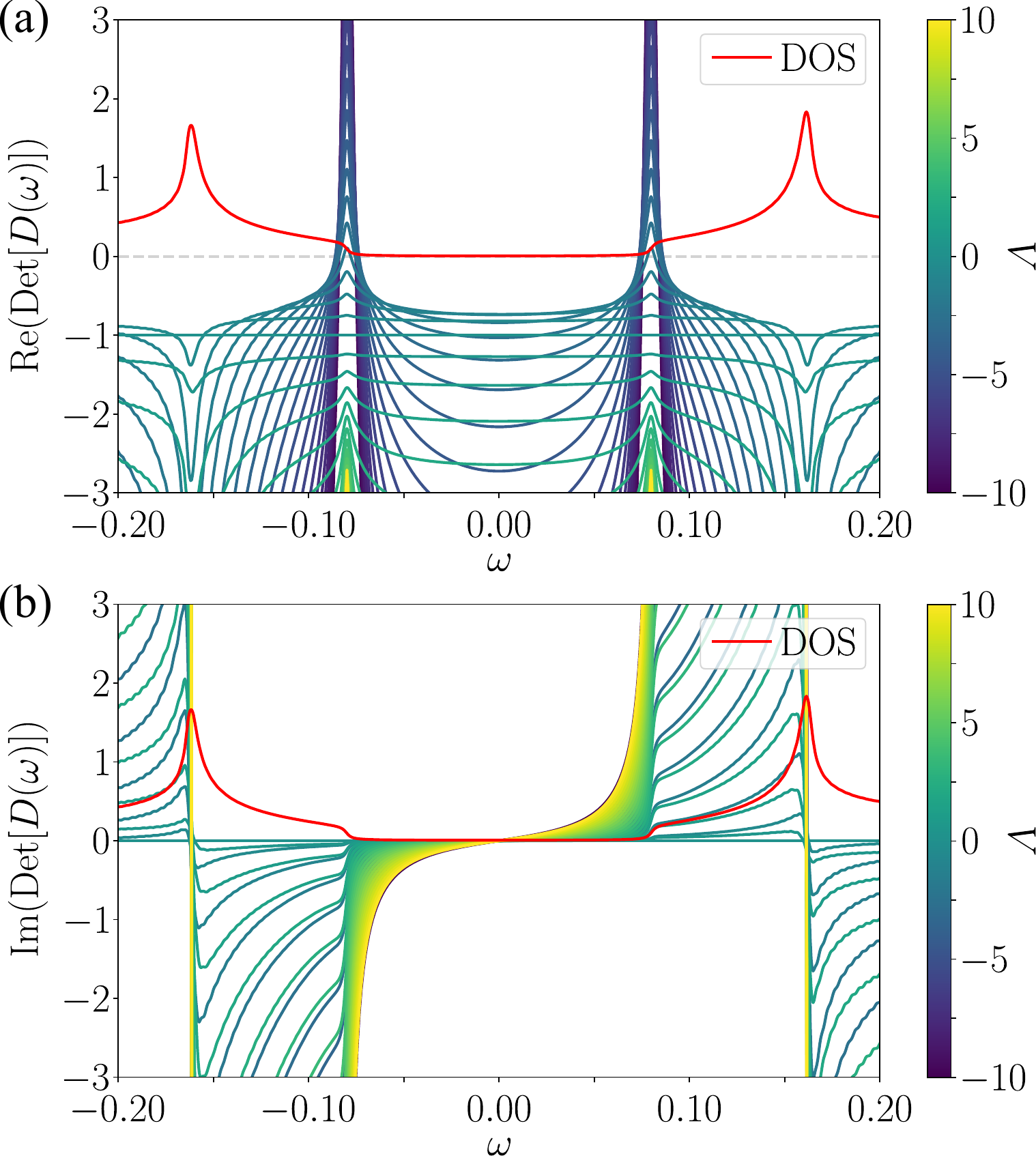}
\caption{\label{fig:dpid_detT}Energy dependence of the real (a) and imaginary (b) parts of $\mbox{det} [\widehat{D}(\omega)]$ as given by Eq.~(\ref{BSconditionp}) for a range of nonmagnetic scattering potentials from $V=-10$ to $V=10$, denoted by the colors, for a singlet $d_{x^2-y^2} + i d_{xy}$ gap at $\mu=0.0$ with the DOS in the clean case shown by the red line. The absence of impurity bound states inside the gapped region is evident from panel (a). We note that the finite $\text{Im} ( \mbox{det} [\widehat{D}(\omega)] )$ inside the hard gap in (b) is due to the finite smearing $\eta=0.002$. Vanishing values of $\eta$ lead to a vanishing imaginary part inside the hard gap.}
\end{figure}

\begin{figure}[t]
\centering
\includegraphics[width=0.9\linewidth]{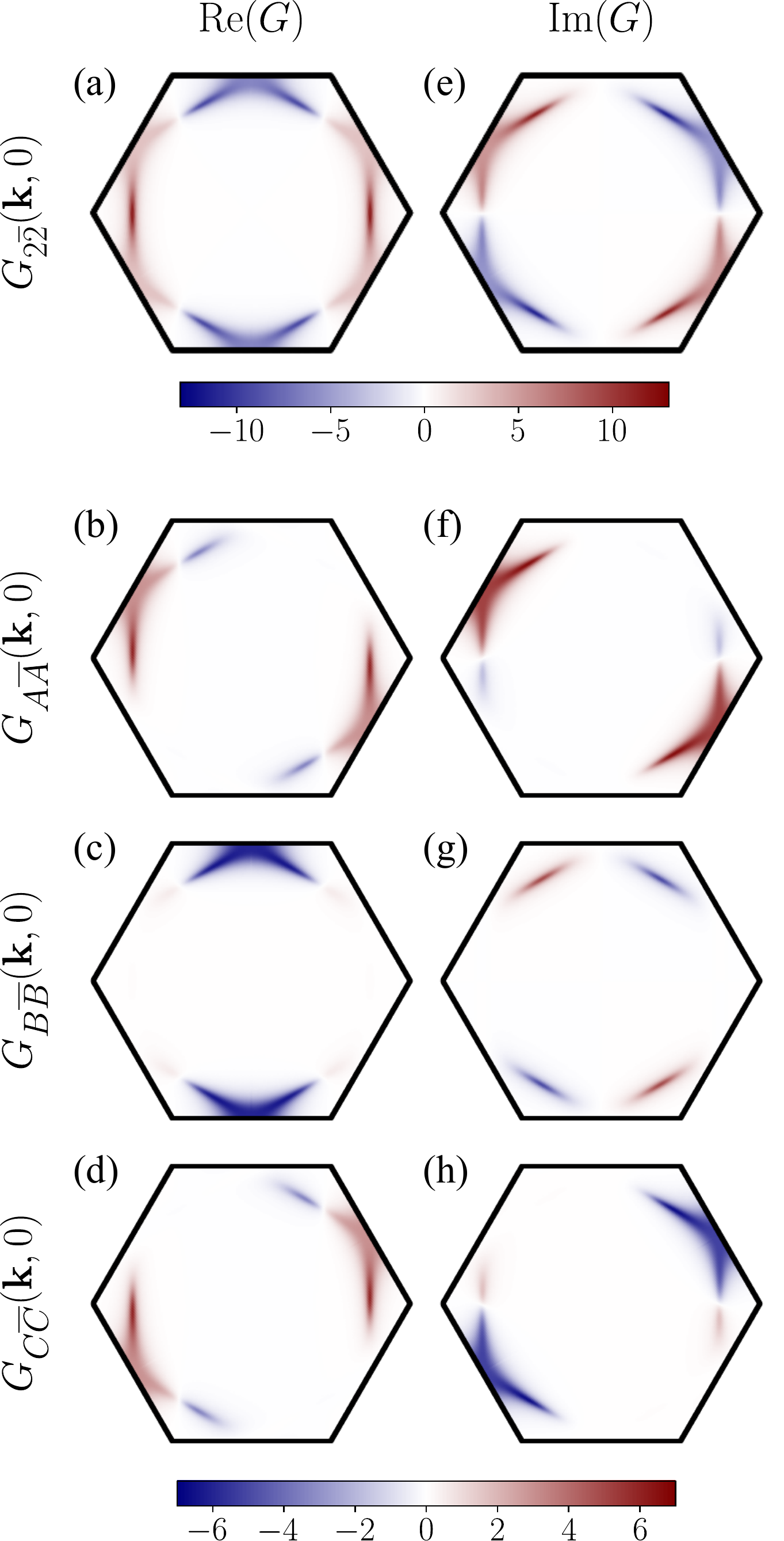}
\caption{\label{fig:d+id_anomalous_Green_mu0_omega0} Momentum dependence of the real (a)--(d) and imaginary (e)--(h) parts of the anomalous free Green function components at $\omega = 0$ for $d_{x^2-y^2}+id_{xy}$ superconductivity at $\mu=0.0$. Panels (a) and (e) display the Green function entries for the second band ($n=2$), reflecting the $d_{x^2-y^2}+id_{xy}$ gap structure at the Fermi surface. The lower six panels refer to the anomalous entries in sublattice space. Importantly, the individual components of the Green function in sublattice space, when integrated over the Brillouin zone, do not vanish, implying that Eq.~\eqref{BSconditionp} may have no real solutions.}
\end{figure}

Turning back to the kagome lattice and the relevant pairing states discussed in the previous section, we find an interesting twist not present in the standard cases summarized above: despite the fact that $\sum_{\mathbf{k}} \Delta_{nm}({\mathbf{k}})=0$ for the spin-singlet order parameters, there are no low-energy in-gap bound state solutions generated by nonmagnetic impurities, i.e. no {\it bona fide} bound state solutions to Eq.~(\ref{BSconditionp}). By contrast, magnetic impurities allow bound state solutions for all superconducting states on the kagome lattice, and will not be further discussed here. 

For the kagome lattice it is not straightforward to obtain analytical expressions for $\sum_{\mathbf{k}} G^{(0)}_{\alpha\beta}(\mathbf{k},\omega)$ and directly analyze Eq.~(\ref{BSconditionp}). However, the absence of low-energy in-gap bound states for spin-singlet order can be illustrated in several ways. In the following, we focus initially on the fully-gapped $d_{x^2-y^2}+id_{xy}$ spin-singlet phase with the homogeneous DOS shown in Fig.~\ref{fig:clean_SC_DOS}(b). Figure~\ref{fig:dpid_detT} shows the real (a) and imaginary (b) parts of the determinant of the $T$-matrix denominator $\widehat{D}(\omega)$ for a range of different scattering potentials. As seen, inside the fully-gapped region the left-hand side of Eq.~(\ref{BSconditionp}) becomes real [Fig.~\ref{fig:dpid_detT}(b)], up to a small imaginary part proportional to the smearing $\eta$, and exhibits no bound state solutions [Fig.~\ref{fig:dpid_detT}(a)]: the low-energy gapped region is protected from any solutions.

What is the origin of the absence of bound states? Figures~\ref{fig:d+id_anomalous_Green_mu0_omega0}(a) and (e) show the gap structure of $d_{x^2-y^2}+id_{xy}$ on the Fermi surface. Clearly both the real and imaginary parts sum to zero. However, for the scattering problem with pointlike defects located at specific sublattice sites, the relevant quantities are the anomalous Green function components in {\it sublattice space}. Figures~\ref{fig:d+id_anomalous_Green_mu0_omega0}(b)--\ref{fig:d+id_anomalous_Green_mu0_omega0}(d) and \ref{fig:d+id_anomalous_Green_mu0_omega0}(f)--\ref{fig:d+id_anomalous_Green_mu0_omega0}(h) display the entries of the off-diagonal sublattice-resolved components of $G_{\alpha \bar \alpha}^{(0)}(\mathbf{k},\omega)$ at $\omega=0$ in  momentum space. As seen, the sublattice weight of the Fermi surface is directly imprinted on the different Green function sublattice components. This implies that the momentum-summed anomalous components, $\sum_{\mathbf{k}} G_{\alpha \bar \alpha}^{(0)}(\mathbf{k},\omega)$, are finite despite the fact that the momentum structure of $d_{x^2-y^2}+id_{xy}$ itself averages to zero over the Fermi surface. In essence, the sublattice weights in conjunction with the even-parity nature of the spin-singlet pairing structure restore the ``standard $s$-wave scenario'' for atomic-scale impurities on the kagome lattice. This is the main finding of the present work.

\begin{figure}[tb]
\centering
\includegraphics[width=0.9\linewidth]{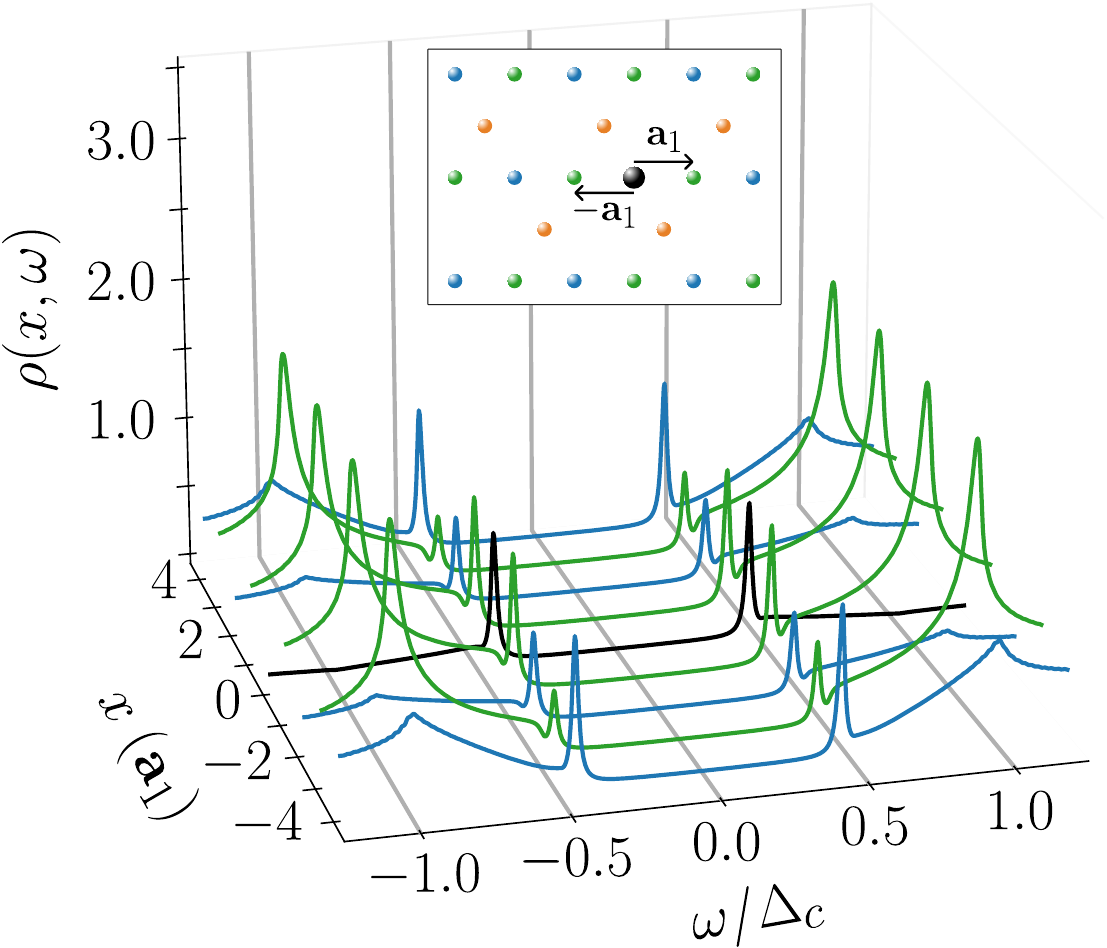}
\caption{\label{fig:dpid_LDOS} LDOS along a cut through the impurity site, shown in the inset, for the case with $d_{x^2-y^2}+id_{xy}$ superconductivity at $\mu = 0.0$ and an impurity potential $V=-4.0$. The different curves correspond to different sites along the cut with the black curve denoting the impurity site, while blue and green curves denote $A$ and $C$ sites, respectively, as shown in the inset. The impurity is located at an $A$ site.}
\end{figure}

In order to further understand the emergence of the above property, we can approximately calculate the relevant components of the anomalous Green function as follows. First, we consider the Green function in band space which (in the absence of interband pairing) becomes block diagonal with the anomalous part 
\begin{equation}
    G_{n\bar n}({\mathbf{k}},\omega)=-\frac{\Delta_n({\mathbf{k}})}{\omega^2-\left(\xi_n(\mathbf{k})-\mu\right)^2-|\Delta_{n}(\mathbf{k})|^2}\,.\label{eq:anomalous_G}
\end{equation}
At low energies, only the band $n^*$ that crosses the Fermi level contributes, and we can ignore anomalous parts of the Green function of other bands. Next, we transform the Green function back to sublattice space (where we want to calculate the local Green function), and use the matrix elements of the unitary transformation $u_{\alpha n}(\mathbf{k})$ to obtain approximately $G_{\alpha\bar\alpha}({\mathbf{k}},\omega)\approx u_{\alpha n^*}(\mathbf{k}) G_{n^*\bar n^*}(\mathbf{k},\omega) u_{n^* \alpha}(-\mathbf{k})$.
Combining this with Eq.~\eqref{eq:anomalous_G}, one sees that the relevant quantity is the product of the sublattice weight and the order parameter $|u_{n^* \alpha}(\mathbf{k})|^2\Delta_{n^*}(\mathbf{k})$, which combined does not average to zero on the Fermi surface even though $\Delta_{n^*}(\mathbf{k})$ does. 

Finally, to complete the discussion of the $d_{x^2-y^2}+id_{xy}$ spin-singlet superconductivity, we show in Fig.~\ref{fig:dpid_LDOS} the LDOS curves in the vicinity of an impurity. In agreement with Fig.~\ref{fig:dpid_detT}, the inner gap is void of any bound state peaks. Interestingly, the gapped region is lined by resonant states, also in agreement with the $T$-matrix analysis from Fig.~\ref{fig:dpid_detT}. 

\begin{figure}[bt]
\centering
\includegraphics[width=\linewidth]{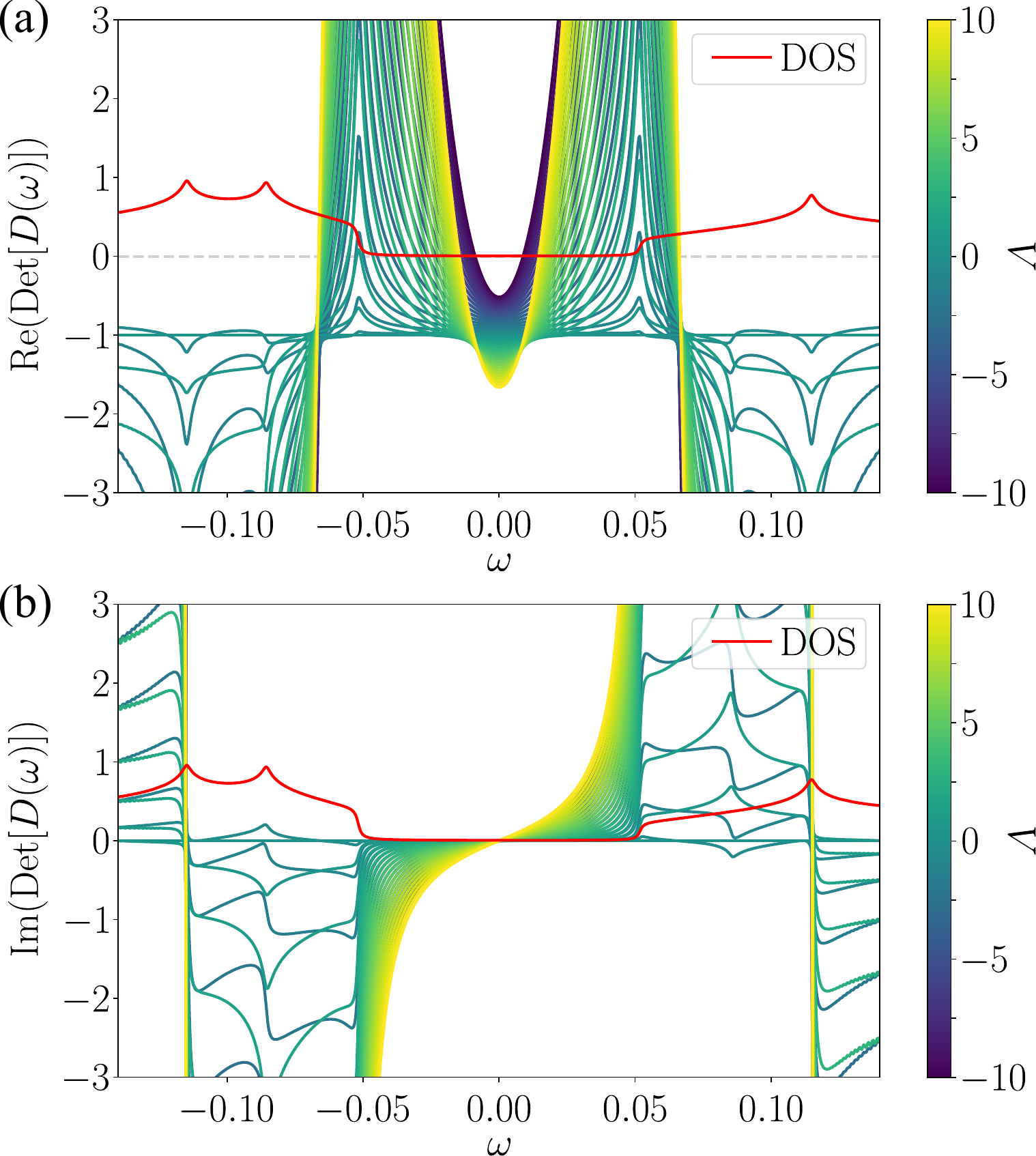}
\caption{\label{fig:ppip_detT}Energy dependence of the real (a) and imaginary (b) parts of $\mbox{det} [\widehat{D}(\omega)]$ as given by Eq.~(\ref{BSconditionp}) for a range of nonmagnetic scattering potentials from $V=-10$ to $V=10$, denoted by the colors, for a triplet $p_x + i p_y$ gap at $\mu=0.08$ with the DOS for the clean case shown by the red line. In this case, the gapped region exhibits impurity bound states, in contrast to the case shown in Fig.~\ref{fig:dpid_detT}. As in Fig.~\ref{fig:dpid_detT}, it is the finite value of $\eta$ that causes a finite imaginary part inside the hard gap in (b).}
\end{figure}

\begin{figure}
\centering
\includegraphics[width=0.9\linewidth]{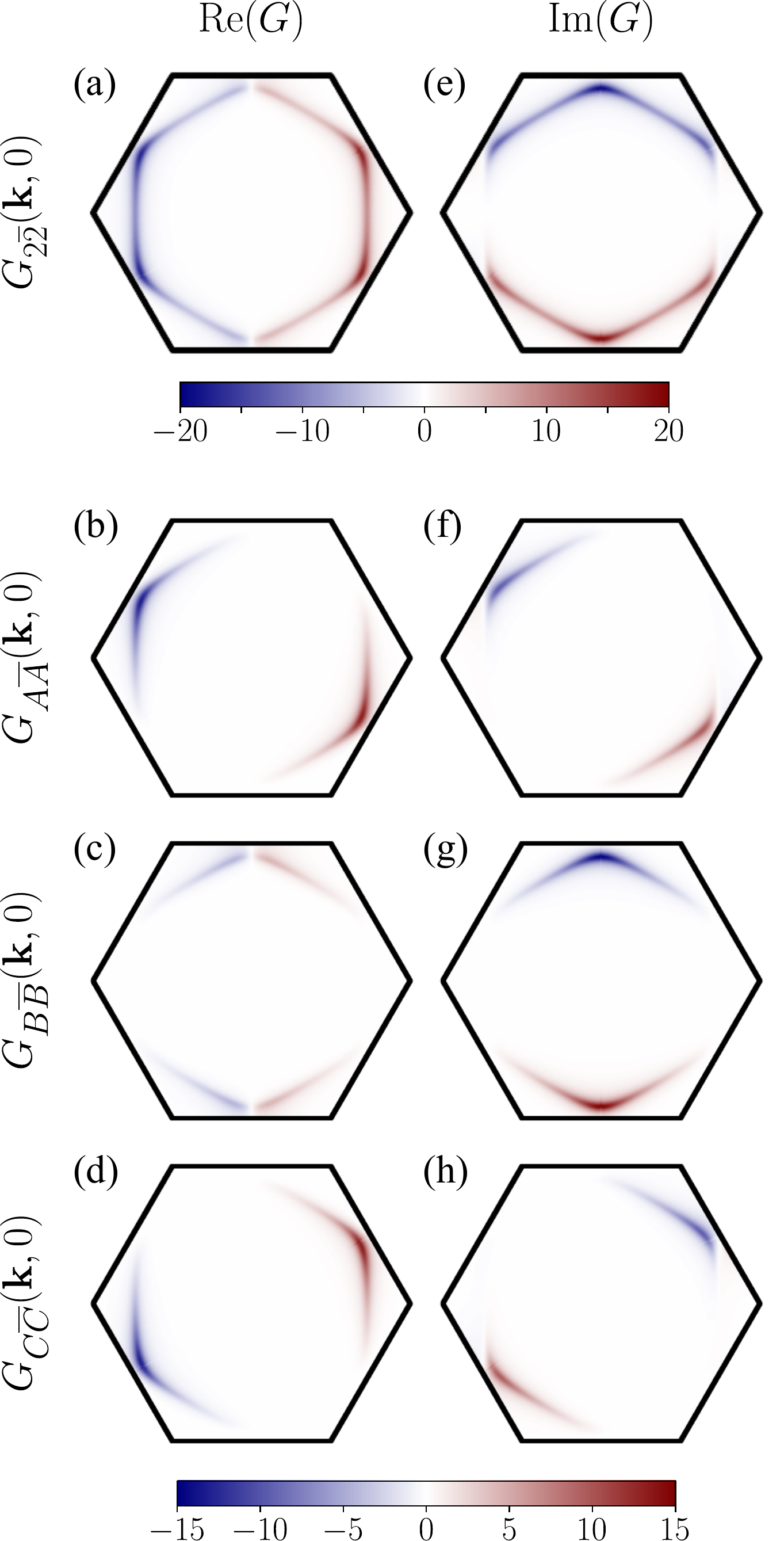}
\caption{\label{fig:ppip_anomalous_Green_functionmu0}Momentum dependence of the real (a-d) and imaginary (e-h) parts of the anomalous free Green function components at $\omega = 0$ for $p_x+ip_{y}$ superconductivity. Panels (a,e) display the Green function entries in band space, simply reflecting the $p_{x}+ip_{y}$ gap structure at the Fermi surface. The lower six panels refer to the anomalous entries in sublattice space. All plots here are for the $p_{x}+ip_{y}$ phase at $\mu = 0.08$. In contrast to the $d_{x^2-y^2} + id_{xy}$ case, the individual components of the sublattice Green function, when integrated over the Brillouin zone, all vanish. As a consequence Eq.~\eqref{BSconditionp} always has real solutions.}
\end{figure}

\begin{figure}
\centering
\includegraphics[width=0.9\linewidth]{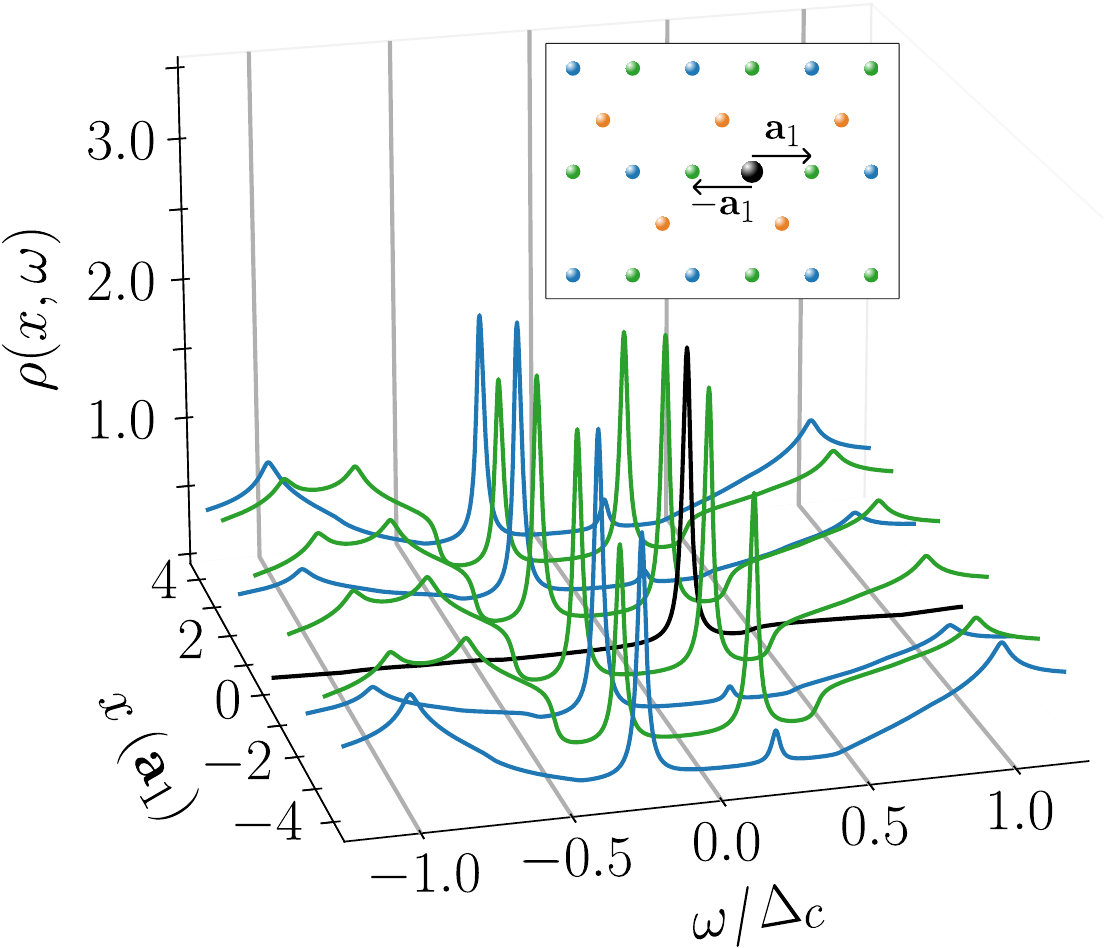}
\caption{\label{fig:ppip_LDOS} LDOS along a cut through the impurity site, shown in the inset, for the case with $p_x + i p_y$ superconductivity at $\mu = 0.08$ and an impurity potential $V=-4.0$. The different curves correspond to different sites along the cut with the black curve denoting the impurity site, while blue and green curves denote $A$ and $C$ sites, respectively, as shown in the inset. The impurity is located on an $A$ site. In contrast to the $d_{x^2-y^2} +id_{xy}$ case shown in Fig.~\ref{fig:dpid_LDOS}, an impurity bound state appears inside the gap for the $p_x + ip_y$ order parameter, and extends along the $\mathbf{a}_1$-direction as shown.} 
\end{figure}

From the above discussion of the mechanism of the protection of spin-singlet superconductivity from nonmagnetic atomic-scale disorder, it is evident that spin-triplet pairing should not enjoy such privilege. This can be verified explicitly by turning to, e.g., the $E_1$ ($p_x+ip_y$) spin-triplet irrep. Figure~\ref{fig:ppip_detT} is equivalent to Fig.~\ref{fig:dpid_detT}, i.e., it shows the real and imaginary parts of $\mbox{det} [\widehat{D}(\omega)]$. Note that the $E_1$ intraband irreps all have nodes at the M points, see Table~\ref{tab:SC_OP}. Hence, to achieve a fully gapped spectrum, we therefore shift the chemical potential to $\mu=0.08$ in this case to move the Fermi surface away from the nodes.  In this setting, we observe that there are bound state solutions at arbitrarily low energies inside the gap. In other words, the odd parity of the triplet order cannot exploit the sublattice weights to ``restore $s$-wave behavior''. Figure~\ref{fig:ppip_anomalous_Green_functionmu0}, which can be compared directly to Fig.~\ref{fig:d+id_anomalous_Green_mu0_omega0}, shows the sublattice-resolved off-diagonal Green function components in momentum space for a spin-triplet $p_x + ip_y$ superconductor. In this case, the sign-changes of the gap function are preserved, also for the relevant combined object  $|u_{n^* \alpha}(\mathbf{k})|^2\Delta_{n^*}(\mathbf{k})$. Therefore, the momentum-summed components vanish, allowing bound state solutions. Finally, in Fig.~\ref{fig:ppip_LDOS} we show the resulting LDOS at the same near-impurity sites as in Fig.~\ref{fig:dpid_LDOS}, all clearly featuring in-gap bound state peaks extended in real space.

\begin{figure*}
\centering
\includegraphics[width=\linewidth]{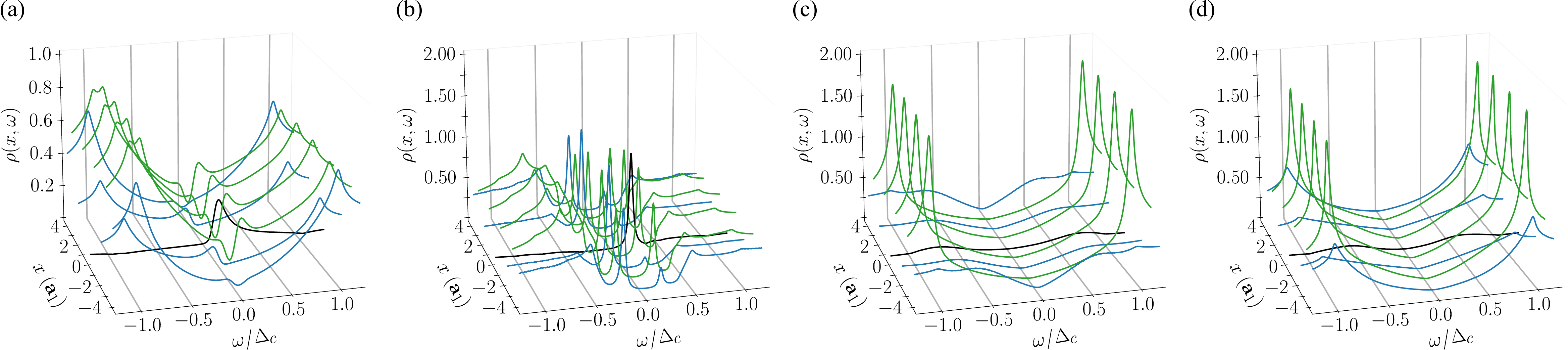}
\caption{\label{fig:LDOSother}LDOS along a horizontal cut (same as Figs.~\ref{fig:dpid_LDOS} and \ref{fig:ppip_LDOS}) near the impurity site (black) for the case of a (a) $B_1$ ($f$-wave), (b) $E^{(1)}_1$ ($p_x$), (c) $E^{(1)}_2$ ($d_{x^2-y^2}$) and (d) $E^{(2)}_2$ ($d_{xy}$) superconducting order parameter with an impurity strength of $V = -4$ in all cases. The spin-singlet cases are obtained with a chemical potential $\mu = 0$, while the triplet cases are calculated with $\mu = 0.08$, i.e. both near the upper van Hove filling. The LDOS for spin-triplet order parameters is visibly affected by the impurity and features low-energy resonant states whereas the singlet $d$-wave order parameters are protected from in-gap resonant states.}
\end{figure*}

Finally we turn to some of the other allowed superconducting order parameters on the kagome lattice and discuss their ability to host bound state solutions from nonmagnetic impurities. First of all, while we demonstrated the relevance (irrelevance) of sublattice weights for fully-gapped spin-singlet (spin-triplet) states of the form $d_{x^2-y^2}+id_{xy}$ ($p_x+ip_y$), we stress that our conclusions remain valid for the individual components, $d_{x^2-y^2}$, $d_{xy}$ ($p_x$, $p_y$), or a real combination of the two. In that case, the LDOS spectrum depends on the sublattice site. In Fig.~\ref{fig:LDOSother} we show cases where the LDOS is nodal and our discussion refers to the absence (presence) of in-gap resonant states for spin-singlet (spin-triplet) order. This is seen from Fig.~\ref{fig:LDOSother} where we show the LDOS similar to Figs.~\ref{fig:dpid_LDOS} and \ref{fig:ppip_LDOS} for (a) $B_1$ ($f$-wave), (b) $E^{(1)}_1$ ($p_x$), (c) $E^{(1)}_2$ ($d_{x^2-y^2}$) and (d) $E^{(2)}_2$ ($d_{xy}$) superconducting order. As discussed above, the crucial property of the gap function is its sign under a parity operation. Therefore, the odd-parity spin-triplet states exhibit in-gap resonant states, whereas the even-parity spin-singlet cases are protected. The $A_2$ spin-singlet even-parity order parameter always allows for in-gap bound states on the kagome lattice, due to the fact that it is odd under $C_2$ rotations around the in-plane axes.

\begin{figure}
\centering
\includegraphics[width=\linewidth]{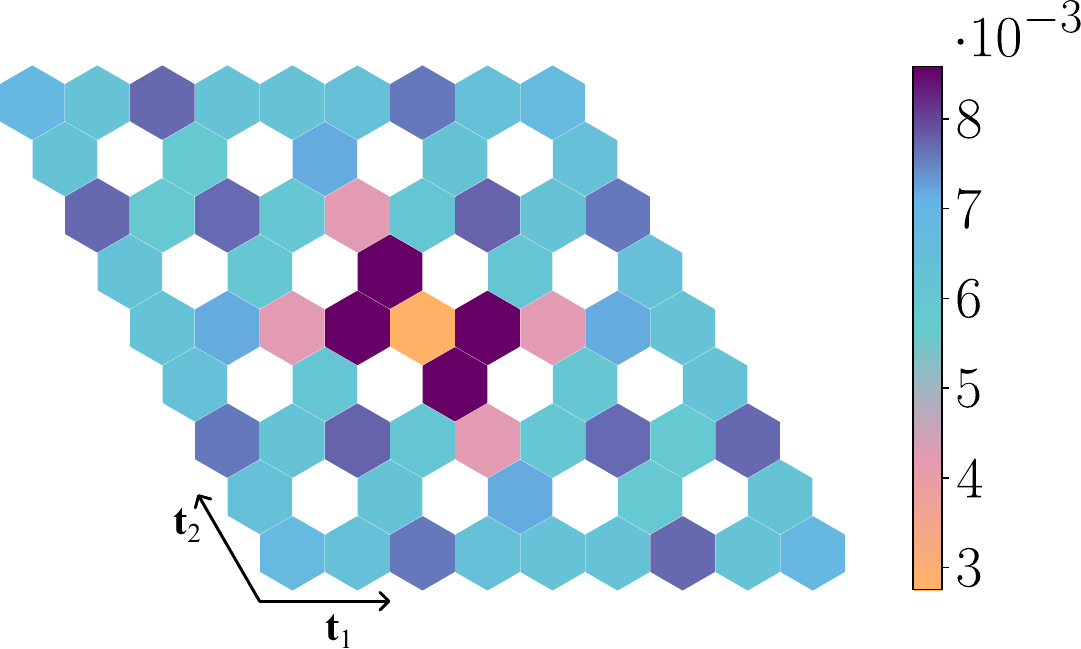}
\caption{\label{fig:QPI_d+id}LDOS $\rho(\mathbf{r},0)$ plotted for a fixed energy $\omega = 0$ with $d_{x^2-y^2}+id_{xy}$ superconductivity and an impurity of strength $V = -4$ located at an A sublattice site. The breaking of several point group symmetries by the impurity position is reflected in the associated LDOS modulations in the vicinity of the impurity.}
\end{figure}

The interplay between sublattice weights and localized impurities may also have important consequences for other probes and measurable quantities. Future studies might explore how competing order may locally dress the impurity potentials. This topic has been studied extensively on other lattices where nonmagnetic impurity bound/resonant states were crucial for generating the induced order~\cite{Tsuchiura2001,ZWang2002,Zhu2002,Chen2004,Andersen2007,Harter2007,Andersen2007,Andersen2010,Schmid_2010,Gastiasoro2013}. Here, we focus on consequences for quasi-particle interference (QPI) measurements from pointlike nonmagnetic defects~\cite{WangLeeQPI,Nunner2006,KreiselQPI}. Figure~\ref{fig:QPI_d+id} shows the LDOS in real-space in the vicinity of a nonmagnetic impurity at the $A$ sublattice site in the case of $d_{x^2-y^2}+id_{xy}$ superconductivity. As seen, the resulting LDOS modulations are highly asymmetric, preserving only $D_2$ symmetry around the impurity site. This is a trivial consequence of the fact that the site-symmetry of the lattice sites is lower than the point group symmetry. Naturally, this anisotropy will also carry over to the QPI images, which will only become approximately symmetric when averaged over large fields of view containing pointlike disorder equally distributed on the three sublattice sites.

\subsection{Disorder-averaged $T_c$-suppression}\label{sec:AG}

The absence of in-gap bound states from nonmagnetic disorder in the case of even-parity unconventional superconducting order on the kagome lattice suggests that atomic-scale disorder may only be weakly pair-breaking. As a consequence, its impact on the superconducting critical temperature $T_c$ may be smaller than naively expected. In order to explore this question, we have calculated the disorder-averaged suppression of the order parameter and $T_c$ within the standard Abrikosov-Gor'kov (AG) framework~\cite{AG_classic}. While this method disallows actual spatial inhomogeneity and neglects local properties when feedback or unusual bandstructure effects are important~\cite{RomerRaising,Gastiasoro2016,Gastiasoro2018}, it provides a simple method to probe and compare the robustness of various superconducting condensates to disorder. 

Within AG-theory, translational invariance is restored upon averaging over random impurity distributions, and the full Green function is
\begin{align}
\label{eq:fullGreen}
    \widehat{G}(\mathbf{k},i\omega_n)^{-1} = \widehat{G}^{(0)}(\mathbf{k},i\omega_n)^{-1} - \widehat{\Sigma}(\mathbf{k},i\omega_n),
\end{align}
where $\widehat{G}^{(0)}(\mathbf{k},i\omega_n)$ is the Green function of the superconducting system and $\widehat{\Sigma}(\mathbf{k},i\omega_n)$ denotes the electronic self-energy. In the Born approximation, the self-energy is
\begin{align}
    \widehat{\Sigma}(i\omega_n) = \frac{nV^2}{N} \sum_{\mathbf{k}} \widehat{h}_{\rm imp} \widehat{G}^{(0)}(\mathbf{k},i\omega_n) \widehat{h}_{\rm imp},
\end{align}
where $N$ is the number of points in the momentum sum, $n$ is the impurity concentration, 
and $h_{\rm imp}$ is the impurity potential as given by Eq.~(\ref{imppotentialp}) without the impurity strength prefactor, i.e. $\widehat{H}_{\rm imp}=V\widehat{h}_{\rm imp}$. The superconducting order parameter is given by
\begin{align}
\label{eq:orderparam}
    \Delta_{\alpha\beta}(\mathbf{k}) = \frac{T V_{\rm SC}}{N} \left(f_{\mathbf{k}}^{\beta\alpha}\right)^{\ast} \sum_{\substack{ \mathbf{k}' \omega_n \\ \gamma\delta}} f^{\gamma\delta}_{\mathbf{k}'} G_{\gamma\bar\delta}(\mathbf{k}',i\omega_n)\,.
\end{align}
The above expression corresponds to an assumed pairing interaction of the form $V^{\alpha\beta\gamma\delta}_{\mathbf{k},\mathbf{k}'} = -V_{\rm SC} \left(f_{\mathbf{k}}^{\beta\alpha}\right)^{\ast}f^{\gamma\delta}_{\mathbf{k'}}$ where $f^{\alpha\beta}_{\mathbf{k}}$ is a form factor transforming as the irrep of our interest, as given by, e.g., the matrices in Sec.~\ref{sec:kagome_SC}. In this manner, a superconducting order parameter transforming as a specific irrep can be stabilized. Thus, we have two inter-dependent equations, Eqs.~(\ref{eq:fullGreen}) and (\ref{eq:orderparam}), which are solved selfconsistently to obtain the order parameter $\Delta(\mathbf{k})$ as a function of temperature, $T$.

\begin{figure}
\centering
\includegraphics[width=0.99\linewidth]{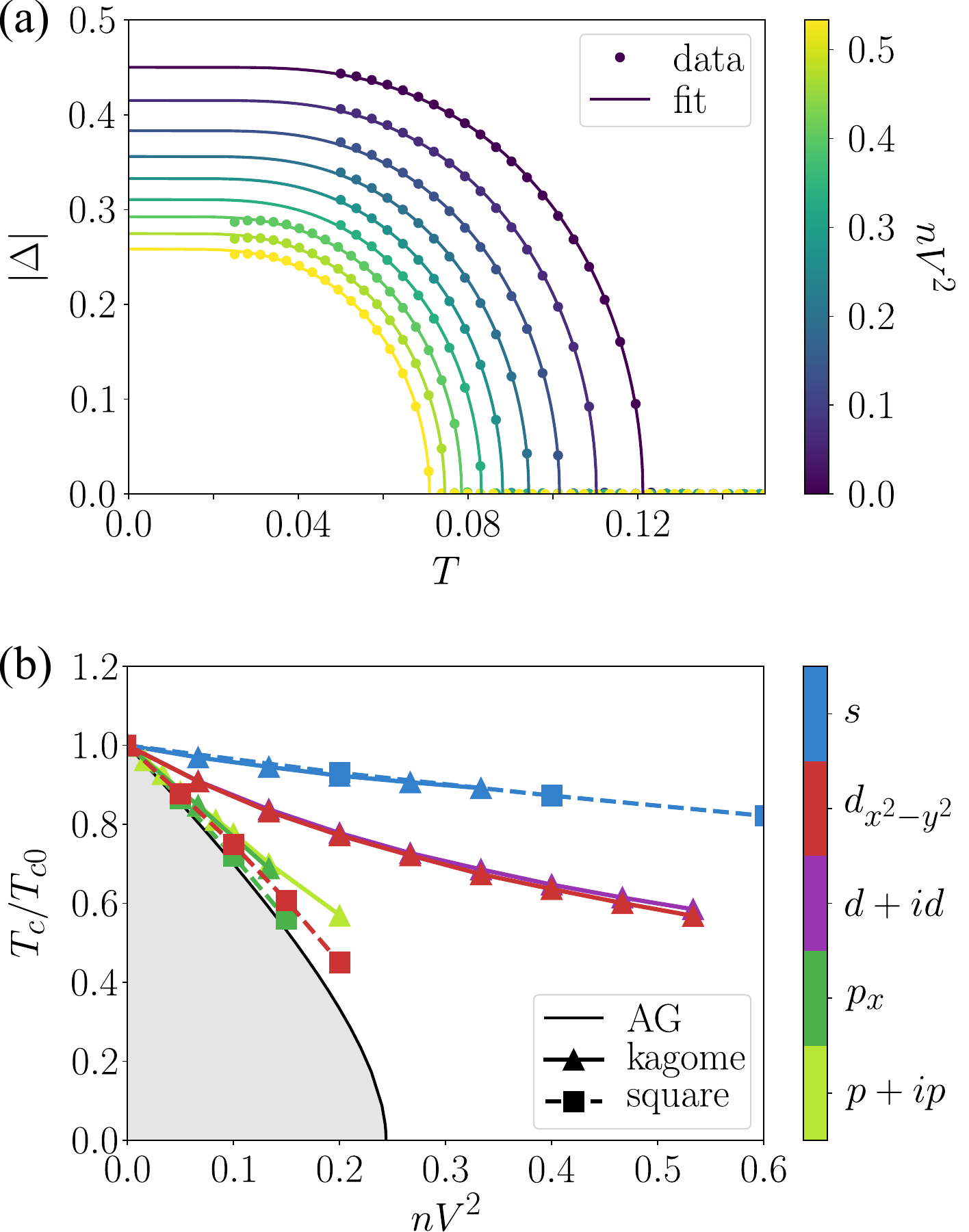}
\caption{\label{fig:AGresults}(a) The temperature dependence of the $d_{x^2-y^2}+id_{xy}$ order parameter $\Delta(T)$ on the kagome lattice for a range of $nV^2$. The AG results (circles) have been fitted with the solutions to the mean-field equation $x = \tanh(x/ T)$ (full lines) to obtain critical temperatures $T_c$. (b) The critical temperatures $T_c/T_{c0}$ as a function of $nV^2$, where $T_{c0} = T_c(nV^2 = 0)$, for an $s$-wave (blue), $d_{x^2-y^2}$ (red), $d+id$ (purple), $p_x$ (dark green) and $p+ip$ (light green) order parameter. Note that the $p_x$ and $p+ip$ results for the kagome lattice nearly coincide. The results are for $\mu = 0$ in the kagome lattice and $\mu = -1$ for the square lattice. In the case of the rotational symmetry breaking order parameters the critical temperatures were calculated with the impurities located on sites A, B and C, respectively, and an average of the three values is plotted. The critical temperatures have been matched in the $nV^2 = 0$ case by self-consistently solving the BdG gap equation. The solid line in panel (b) shows the $T_c$-suppression from the universal AG curve~\cite{AG_classic,RomerRaising}.}
\end{figure}

In Fig.~\ref{fig:AGresults}(a) we show the temperature dependence of the superconducting gap,
\begin{equation}
\Delta \equiv \frac{T V_{\rm SC}}{N} \sum_{\substack{ \mathbf{k} \,  \omega_n \\ \gamma\delta}} f^{\gamma\delta}_{\mathbf{k}} G_{\gamma\bar\delta}(\mathbf{k},i\omega_n)\,,
\end{equation}
in the $d_{x^2-y^2}+id_{xy}$ case for different impurity concentrations. Note that this deviates from Eq.~\eqref{eq:orderparam} by the form factor, $(f^{\beta\alpha}_{\mathbf{k}})^{\ast}$, which only serves to ensure the correct sublattice and momentum dependence of the order parameter. To determine the critical temperature $T_c$, we fit the solutions to the mean-field equation $x = \tanh(x/ T)$, where $x$ is the mean-field order parameter, to our results. This yields the solid lines in Fig.~\ref{fig:AGresults}(a). In Fig.~\ref{fig:AGresults}(b), we plot $T_c$, determined from fitting to the AG results in Fig.~\ref{fig:AGresults}(a), as a function of impurity concentration $nV^2$ for different superconducting order parameters (colored lines) and square and kagome lattices, corresponding to dashed and full lines, respectively. The $d_{x^2-y^2}+id_{xy}$ case shown in Fig.~\ref{fig:AGresults}(a) leads to the full purple curve in Fig.~\ref{fig:AGresults}(b). The same suppression rate is obtained for $d_{x^2-y^2}$ superconductivity on the kagome lattice, as shown by the full red curve in Fig.~\ref{fig:AGresults}(b). In this case, we have averaged over impurities on the $A$, $B$, and $C$ sublattices and the horizontal axis is rescaled taking into account that the impurity potential only acts on $1/3$ of the sites in the kagome lattice. The latter applies to all kagome results in Fig.~\ref{fig:AGresults}(b). This is to be contrasted with $d_{x^2-y^2}$ superconductivity on the square lattice, which is shown by the red dashed line in Fig.~\ref{fig:AGresults}(b), which follows the standard AG $T_c$-suppression curve~\cite{AG_classic,RomerRaising}. The AG-curve in Fig.~\ref{fig:AGresults}(b) corresponds to a case with $\rho(0)=0.14$, the DOS at $\mu=-1$ for the square lattice, and is also representative for the kagome DOS near $\mu=0$ (averaged over the gap energy scale). Remarkably, the $T_c$-suppression of even-parity unconventional superconductivity on the kagome lattice is qualitatively distinct from the $T_c$ suppression of unconventional superconductivity on the square lattice. We ascribe the origin of the reduced pair-breaking on the kagome lattice to the interplay between sublattice weights and localized impurities, as discussed above. In Fig.~\ref{fig:AGresults}(b) we additionally compare the $T_c$-suppression to standard $s$-wave order and triplet order on the square and kagome lattices. As seen, standard $s$-wave order exhibits the expected slow dependence on impurity concentration in agreement with Anderson's theorem, whereas all triplet orders remain fragile to nonmagnetic disorder, also on the kagome lattice since it is not sublattice-protected.

\section{Discussion and conclusions}\label{sec:discussion}

In this work, after a general discussion of the allowed homogeneous superconducting order parameters on the kagome lattice, we have focused on their properties in the presence of atomic-scale defects. By picking specific sublattice sites, such isolated impurities limit scattering between Bloch eigenstates to selected regions of the Fermi surface. This leads to an important difference between even- and odd-parity superconducting order parameters in terms of the ability of nonmagnetic impurities to generate bound states and break Cooper pairs. This fundamental distinction between the response to nonmagnetic disorder between spin-singlet and spin-triplet order is unlike, e.g., the square lattice~\cite{WangImpurity2004,Balatsky_review,Andersen2006_Andreev} or earlier results relevant for graphene on the honeycomb lattice\cite{Pellegrino2010,LothmanDefects2014} where both such superconducting condensates support impurity bound or resonant states, and become fragile to nonmagnetic disorder. For even-parity superconductivity on the kagome lattice, pointlike disorder studies are not phase-sensitive probes of superconductivity.

We stress that the absence of in-gap bound states in even-parity superconductors from nonmagnetic disorder is a generic property of the kagome lattice, and not tied to fine-tuned parameters of, e.g., the Fermi surface topology or the amplitude of the impurity potential. We have focused on the upper van Hove filling due to its relevance for the $A$V$_3$Sb$_5$ compounds and, at that filling, the protected gap region is very pronounced. However, the Fermi surface states exhibit significant momentum-dependent variation of the sublattice weight also at other fillings, thus leading to qualitatively similar effects.

\begin{figure}[tb]
\centering
\includegraphics[width=0.9\linewidth]{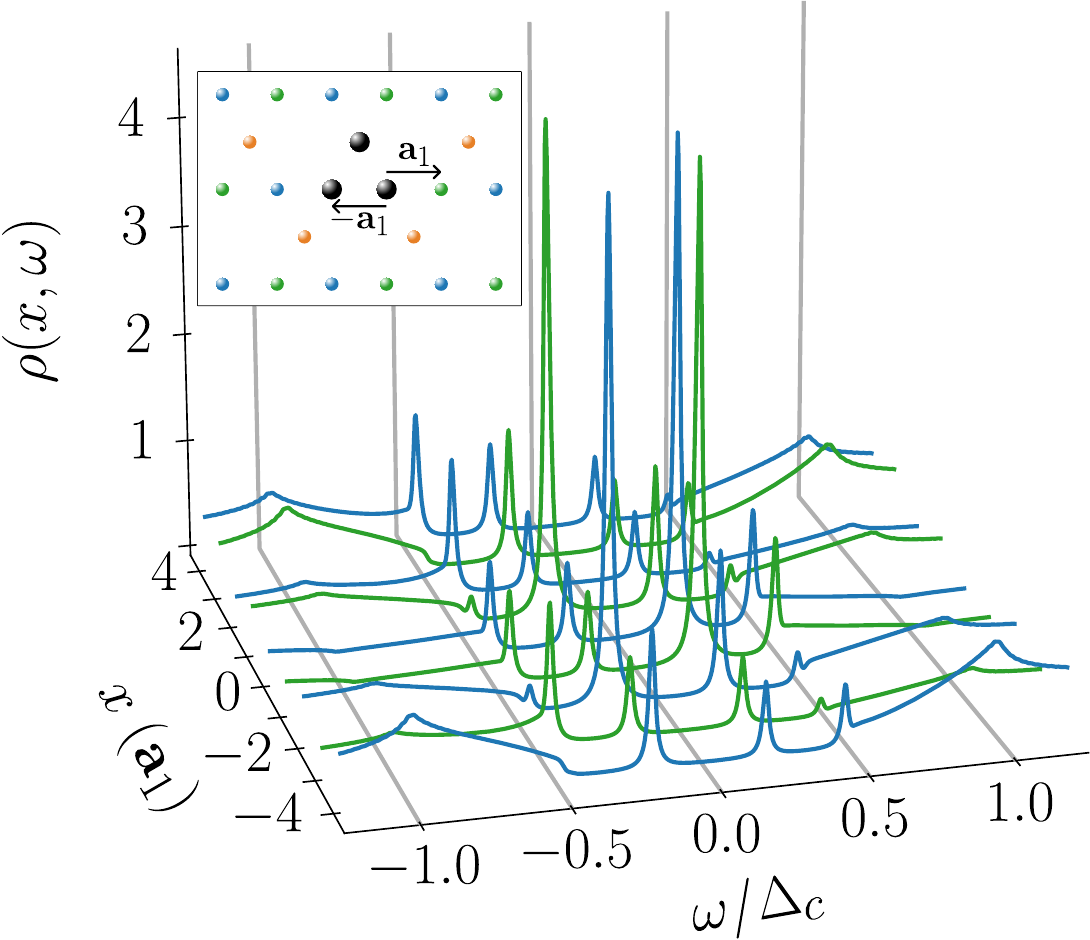}
\caption{\label{fig:dpid_LDOS_diagimp}LDOS near an impurity potential extended equally to the $A$, $B$ and $C$ sublattice sites, as shown in the inset, for a $d_{x^2-y^2}+id_{xy}$ superconductor at $\mu = 0.0$ and an impurity potential $V=-2.0$. In the present case impurity bound states emerge inside the gap and the in-gap bound state position is tuned by $V$.}
\end{figure}

As mentioned above, the amplitude of the pointlike impurity potential is not important. Likewise, whether the impurity resides on sublattice site $A$, $B$ or $C$, is also not important. These conclusions are based on single-impurity properties and thus valid in the dilute disorder limit. When disorder cross-talk becomes important or if the impurities correspond to spatially extended objects, the disorder-response is altered. For example, if the disorder potential extends equally to all three sublattice sites, impurity scattering probes the whole Fermi surface and bound states may get generated, also for an even-parity spin-singlet sign-changing superconducting order. This is illustrated in Fig.~\ref{fig:dpid_LDOS_diagimp} where we show that there are indeed in-gap bound state solutions for an extended triangular impurity. Since interstitials or impurities off the kagome plane are expected to produce extended disorder potentials, such perturbations can generate in-gap bound states. However, in such cases there is the additional complication of the strengths of the potentials generated by more extended defects and whether they are strong enough to produce low-energy bound states. In particular, if the disorder potentials arise from out-of-plane defects or interstitials, the resulting scattering potentials in the kagome planes may simply be too weak to generate in-gap bound states.

An outstanding question relates to the relevance of the mechanism described in this work of how sublattice weights can ``restore $s$-wave behavior'' for the $A$V$_3$Sb$_5$ superconductors. These materials exhibit a significantly more complex Fermi surface than the NN tight-binding band utilized in this work~\cite{Ortiz2020CsV3Sb5}. Thus, for the current mechanism to be active in $A$V$_3$Sb$_5$, superconductivity in these compounds should primarily be present on the vanadium $d$-orbital bands featuring significant sublattice weight variation in momentum space. Furthermore, these materials exhibit CDW order in addition to superconductivity. However, the CDW order leads to only a weak electronic reconstruction~\cite{Tjernberg2022,Kato2022} and it 
should therefore not be important for the current discussion.  Lastly, as we have stressed, the main source of disorder has to be pointlike defects located on specific sublattice sites. This seems to be relevant, for example, to the STM experiments reported in Ref.~\onlinecite{Xu2021Multiband} finding no impurity-bound states near vanadium vacancy defects. Likewise, the irradiation studies of the penetration depth in Ref.~\onlinecite{RoppongiEA22} introduce mainly pointlike defects. Within our scenario, the in-plane vanadium defects are not phase sensitive to the superconducting gap structure, whereas in-plane Sb or out-of-plane Sb and Cs defects might be. The latter depends on their strength on the vanadium sites in the kagome planes remaining after screening. 

The understanding of slow $T_c$-suppression rates has also been recently discussed in the context of e.g. PdTe$_2$ and Cu$_x$Bi$_2$Se$_3$~\cite{Michaeli2012,Timmons2020,Andersen2020,Cavanagh2020,Zinkl2022}. Related theoretical studies have formulated a “generalized Anderson theorem”, clarifying under what circumstances the nature of the disorder is benign for the particular superconducting state under consideration~\cite{Scheurer2016_1000056491,Timmons2020,Andersen2020}. In the present case, the weak $T_c$-suppression rate cannot be straightforwardly explained from the generalized Anderson theorem since the gap structure and the disorder potential do not fulfil the basic requirements of the theorem. This is mainly due to the existence of interband components of the superconducting gap structure. Within the formulation of Ref.~\onlinecite{Andersen2020}, an additional distinction of the current work is the presence of momentum dependence of the superconducting gap structure in sublattice space. Thus, the current “robustness” of superconductivity is not quantitatively in line with Anderson’s theorem, in agreement with the weaker robustness seen in Fig.~\ref{fig:AGresults}. Finally, we note an interesting study of robust superfluidity in the polar phase of $^3$He in the presence of oriented columnar defects~\cite{Kamppinen2023}. In that case the resulting anisotropic scattering of the columnar disorder is blind to the putative nodal line of the gap, not unlike the present case where the pointlike defects are also highly anisotropic scatterers in momentum space.

In this manuscript we have mainly studied unconventional superconductivity on the kagome lattice. Naturally, if it turns out that the $A$V$_3$Sb$_5$ materials are conventional superconductors with standard non-sign-changing $s$-wave superconductivity, there is nothing for the sublattice weights ``to restore'' and the robustness to nonmagnetic disorder follows from Anderson's theorem~\cite{ANDERSON195926}. The main point of the present paper is to point out that even unconventional sign-changing spin-singlet superconducting orders on the kagome lattice exhibit an intrinsic robustness to nonmagnetic disorder. 

\begin{acknowledgments}
We acknowledge fruitful discussions with F. Ferrari, M. Scheurer, and T. Shibauchi. This project (M.H.C.) has received funding from the European Union’s Horizon 2020 research and innovation programme under the Marie Sklodowska-Curie grant agreement No 101024210. A.K. acknowledges support by the Danish National Committee for Research Infrastructure (NUFI) through the ESS-Lighthouse Q-MAT.
\end{acknowledgments}

\appendix

\section{Pairing matrices}\label{app:PairingMatrices}

Here we include the explicit form of all the form factor matrices as defined in Eqs.~\eqref{eq:f_os_singlet}, \eqref{eq:f_nn_singlet}, and \eqref{eq:f_nn_triplet} for both singlet and triplet order parameters up to nearest neighbor. For the $A_1$ form factor we find
\begin{align}
    f^{S}_{{\rm OS},A_{1}} &= \frac{1}{\sqrt{3}}\begin{pmatrix}
        1 & 0 & 0 \\
        0 & 1 & 0 \\
        0 & 0 & 1
    \end{pmatrix}\,, \\
    f^{S}_{{\rm NN},A_{1}} &= \frac{1}{\sqrt{6}} \begin{pmatrix}
        0 & \cos k_3 & \cos k_1 \\
        \cos k_3 & 0 & \cos k_2 \\
        \cos k_1 & \cos k_2 & 0
    \end{pmatrix}\,,
\end{align}
while for the singlet $B_2$ we have:
\begin{align}
    f^S_{{\rm NN},B_2} = \frac{i}{\sqrt{6}}\begin{pmatrix}
        0 & -\sin k_3 & \sin k_1 \\
        \sin k_3 & 0 & \sin k_2 \\
        -\sin k_1 & -\sin k_2 & 0
    \end{pmatrix}\,.
\end{align}
Due to the sublattice degree of freedom, there is both a singlet and a triplet order parameter with $E_1$ character. The singlet form factors are
\begin{align}
    f^S_{{\rm NN},E_1^{(1)}} &= \frac{i}{2}\begin{pmatrix}
        0 & \sin k_3 & 0 \\
        -\sin k_3 & 0 & \sin k_2 \\
        0 & - \sin k_2 & 0
    \end{pmatrix}\,, \\
    f^S_{{\rm NN},E_1^{(2)}} &= \frac{i}{\sqrt{12}}\begin{pmatrix}
        0 & \sin k_3 & 2 \sin k_1 \\
        -\sin k_3 & 0 & -\sin k_2 \\
        -2\sin k_1 & \sin k_2 & 0
    \end{pmatrix}\,.
\end{align}
The singlet $E_2$ form factor has both OS and NN contributions:
\begin{align}
    f^S_{{\rm OS},E_2^{(1)}} &= \frac{1}{\sqrt{6}} \begin{pmatrix}
        +1 & 0 & 0 \\
        0 & -2 & 0 \\
        0 & 0 & +1
    \end{pmatrix}\,, \\
    f^S_{{\rm OS},E_2^{(2)}} &= \frac{1}{\sqrt{2}} \begin{pmatrix}
        +1 & 0 & 0 \\
        0 & 0 & 0 \\
        0 & 0 & -1
    \end{pmatrix}\,, \\
    f^S_{{\rm NN},E_2^{(1)}} &= \frac{1}{\sqrt{12}} \begin{pmatrix}
        0 & -\cos k_3 & 2 \cos k_1 \\
        - \cos k_3 & 0 & -\cos k_2 \\
        2 \cos k_1 & -\cos k_2 & 0
    \end{pmatrix}\,, \\
    f^S_{{\rm NN},E_2^{(2)}} &=\frac{1}{2} \begin{pmatrix}
        0 & \cos k_3 & 0 \\
        \cos k_3 & 0 & -\cos k_2 \\
        0 & -\cos k_2 & 0
    \end{pmatrix}
\end{align}

The triplet orders occur at nearest-neighbor and beyond. For the $A_2$ form factor, we find
\begin{align}
    f^{T}_{{\rm NN},A_{2}} = \frac{1}{\sqrt{6}} \begin{pmatrix}
        0 & -\cos k_3 & \cos k_1 \\
        \cos k_3 & 0 & -\cos k_2 \\
        -\cos k_1 & \cos k_2 & 0
    \end{pmatrix}\,,
\end{align}
and for $B_1$:
\begin{align}
    f^{T}_{{\rm NN},B_{1}} = \frac{i}{\sqrt{6}} \begin{pmatrix}
        0 & -\sin k_3 & -\sin k_1 \\
        -\sin k_3 & 0 & \sin k_2 \\
        -\sin k_1 & \sin k_2 & 0
    \end{pmatrix}\,.
\end{align}
The $E_1$ triplet form factor takes the form
\begin{align}
    f^T_{{\rm NN},E_1^{(1)}} &=\frac{i}{\sqrt{12}}\begin{pmatrix}
        0 & \sin k_3 & -2 \sin k_1 \\
        \sin k_3 & 0 & -\sin k_2 \\
        -2\sin k_1 & -\sin k_2 & 0
    \end{pmatrix}\,, \\
    f^T_{{\rm NN},E_1^{(2)}} &= \frac{i}{2}\begin{pmatrix}
        0 & \sin k_3 & 0 \\
        \sin k_3 & 0 & \sin k_2 \\
        0 &  \sin k_2 & 0
         \end{pmatrix}\,,
\end{align}
while we find the $E_2$ triplet form factor to be
\begin{align}
    f^T_{{\rm NN},E_2^{(1)}} &=\frac{1}{2} \begin{pmatrix}
        0 & -\cos k_3 & 0 \\
        \cos k_3 & 0 & \cos k_2 \\
        0 & -\cos k_2 & 0
    \end{pmatrix}\,, \\
    f^T_{{\rm NN},E_2^{(2)}} &= \frac{1}{\sqrt{12}} \begin{pmatrix}
        0 & \cos k_3 & 2 \cos k_1 \\
        - \cos k_3 & 0 & \cos k_2 \\
        -2 \cos k_1 & -\cos k_2 & 0
    \end{pmatrix}\,.
\end{align}

\bibliography{Kagome}
\end{document}